\documentclass[12pt,a4paper]{article}
\usepackage[utf8]{inputenc}

\usepackage{jheppub}

\makeatletter
\renewcommand*\env@matrix[1][*\c@MaxMatrixCols c]{%
  \hskip -\arraycolsep
  \let\@ifnextchar\new@ifnextchar
  \array{#1}}
\makeatother
 
\usepackage{amsmath}
\usepackage{amssymb}
\usepackage{braket}
\usepackage{array}
\newcolumntype{M}[1]{>{\centering\arraybackslash}m{#1}}

\usepackage{xcolor}
\usepackage{graphicx}
\usepackage{subcaption}

\usepackage[compat=1.1.0]{tikz-feynman}
\usepackage{tikz}

\usepackage[compat=1.1.0]{tikz-feynman}
\definecolor{airforceblue}{rgb}{0.36, 0.54, 0.66}
\definecolor{blue(ncs)}{rgb}{0.0, 0.53, 0.74}
\definecolor{caribbeangreen}{rgb}{0.0, 0.8, 0.6}
\definecolor{green(ncs)}{rgb}{0.0, 0.62, 0.42}
\definecolor{myblue}{RGB}{135, 189, 1}
\definecolor{darkgreen}{RGB}{1, 130, 1}

\def\b#1{{\color{blue(ncs)}#1}}
\def\r#1{{\color{red}#1}}
\def\g#1{{\color{gray}#1}}
\def\gn#1{{\color{darkgreen}#1}}

\newcommand{\mtabref}[1]{Table~\ref{#1}}
\newcommand{\mfigref}[1]{Figure~\ref{#1}}
\newcommand{\msecref}[1]{section~\ref{#1}}

\newcommand{\ghost}[0]{$\sqrt{\text{dilaton}}$}
\newcommand{\ghosts}[0]{$\sqrt{\text{dilaton}}$s}

\def\eqn#1{eq.~\eqref{#1}}
\def\eqns#1#2{eqs.~\eqref{#1} and~\eqref{#2}}
\def\eqnss#1#2#3{eqs.~\eqref{#1}, \eqref{#2} and~\eqref{#3}}

\def\rcite#1{ref.~\cite{#1}}
\def\rcites#1{refs.~\cite{#1}}

\def\be{\begin{equation}}
\def\ee{\end{equation}}
\def\bea{\begin{eqnarray}}
\def\eea{\end{eqnarray}}
\def\beal{\begin{equation}\begin{aligned}}
\def\eeal{\end{aligned}\end{equation}}
\def\nn{\nonumber}

\def\fD{f_{\pm}}
\def\fDp{f_{+}}
\def\fDm{f_{-}}

\title{Gleaning gravitational amplitudes -- a double copy for canceling dilatons}

\author[1,2]{Henrik Johansson,}
\author[1,2]{Ingrid A. Vazquez-Holm}

\affiliation[1]{Department of Physics and Astronomy, Uppsala University, Box 516, 75120 Uppsala, Sweden}
\affiliation[2]{Nordita, Stockholm University and KTH Royal Institute of Technology, Hannes Alfv\'ens v\"ag 12, 10691 Stockholm, Sweden}

\emailAdd{henrik.johansson@physics.uu.se} \emailAdd{ingrid.holm@physics.uu.se}

\preprint{UUITP-2/25 
}

\abstract{
Scattering amplitudes in general relativity can be conveniently computed using the double copy, which relates them to Yang-Mills amplitudes. However, unwanted dilatons are sourced by massive scalar matter, which must be removed from the double copy in order to match the long range gravitational interactions. In this paper, we study how to automatically cancel out the dilatons by finding a suitable double-copy prescription in terms of gauge-theory fields, effectively treating the new contributions as ghosts that subtract out the unwanted states. At tree level, we find that an asymmetric double copy can reproduce the dilaton graphs in general dimension, which we explicitly verify up to six external massive scalars. Considering a one-loop four-point example, the same asymmetric double copy needs to be supplemented by the subtraction of bubble graphs that originate both from the axion and a residual dilaton term.  
}

\begin{document}

\maketitle

\pagebreak 

\section{Introduction}

Gauge and gravity theories exhibit many useful mathematical structures that are visible only after computing scattering amplitudes. The duality between color and kinematics~\cite{Bern:2008qj}, and the gravitational double copy~\cite{Bern:2010ue}, are two notable examples of such structures that provide computational efficiency. The duality implies that kinematic numerators of cubic diagrams can be made to exhibit the same Lie-algebra relations as the corresponding color factors, and the double copy recycles the numerators giving gravitational amplitudes from two copies of gauge theories. See recent reviews on the color-kinematics duality and double copy~\cite{Bern:2019prr,Bern:2022wqg,Adamo:2022dcm,McLoughlin:2022ljp,Berkovits:2022ivl,Bern:2022jnl,Mafra:2022wml}.

For pure Yang-Mills (YM) theory, or any duality-satisfying massless gauge theory with only adjoint matter, the color-kinematics duality implies the existence of Bern-Carrasco-Johansson (BCJ) relations between partial amplitudes~\cite{Bern:2008qj,Stieberger:2009hq,BjerrumBohr:2009rd,Feng:2010my,BjerrumBohr:2010hn}. Both the duality and amplitude relations were first established in the context of pure YM theory~\cite{Bern:2008qj} and related supersymmetric cousins~\cite{Bern:2010ue,Bern:2010yg,Stieberger:2009hq,BjerrumBohr:2009rd}. Over the years, similar or identical color-kinematics dualities have been found in a multitude of interesting gauge theories, encompassing those with matter representations~\cite{Chiodaroli:2013upa,Johansson:2014zca,Chiodaroli:2014xia,Johansson:2015oia,Chiodaroli:2015rdg,Chiodaroli:2018dbu,Johansson:2019dnu,Bautista:2019evw,Plefka:2019wyg}, higher-derivative interactions~\cite{Broedel:2012rc,Johansson:2017srf,Johansson:2018ues,Azevedo:2018dgo, Garozzo:2018uzj,Carrasco:2021ptp,Chi:2021mio,Menezes:2021dyp, Bonnefoy:2021qgu,Carrasco:2022lbm,Carrasco:2022jxn,Carrasco:2022sck}, massive or spontaneously broken gauge theories~\cite{Naculich:2014naa,Naculich:2015zha,Johansson:2015oia,Chiodaroli:2015rdg,Johansson:2019dnu,Momeni:2020vvr, Johnson:2020pny, Moynihan:2020ejh,Momeni:2020hmc,Gonzalez:2021bes,Moynihan:2021rwh, Gonzalez:2021ztm,Chiodaroli:2022ssi,Li:2021yfk,Gonzalez:2022mpa,Li:2022rel,Emond:2022uaf,Engelbrecht:2022aao}, or with a Chern-Simons field~\cite{Bargheer:2012gv,Huang:2012wr,Huang:2013kca, Sivaramakrishnan:2014bpa,Ben-Shahar:2021zww}.
Venturing beyond gauge theories, the duality extends to some well-known scalar effective field theories~\cite{Chen:2013fya,Cheung:2016prv,Carrasco:2016ldy,Mafra:2016mcc,Carrasco:2016ygv,Low:2019wuv,Cheung:2020qxc,Rodina:2021isd,deNeeling:2022tsu, Cheung:2022vnd}. Color-kinematics duality has natural extensions to loop-level amplitudes~\cite{Bern:2010ue, Carrasco:2011mn, Bern:2012uf, Boels:2013bi, Bjerrum-Bohr:2013iza, Bern:2013yya, Nohle:2013bfa, Mogull:2015adi, Mafra:2015mja, He:2015wgf,Johansson:2017bfl, Hohenegger:2017kqy, Mafra:2017ioj, Faller:2018vdz, Kalin:2018thp, Ben-Shahar:2018uie, Duhr:2019ywc, Geyer:2019hnn, Edison:2020uzf, Casali:2020knc, DHoker:2020prr, Carrasco:2020ywq, Bridges:2021ebs,Guillen:2021mwp,Carrasco:2021bmu,Porkert:2022efy,Edison:2022smn,Edison:2022jln}, form factors~\cite{Boels:2012ew,Yang:2016ear,Boels:2017ftb,Lin:2020dyj,Lin:2021kht,Lin:2021pne,Lin:2021lqo, Lin:2021qol,Chen:2022nei,Li:2022tir}, and even to correlators in curved space~\cite{Adamo:2017nia,Farrow:2018yni,Adamo:2018mpq,Lipstein:2019mpu, Prabhu:2020avf, Armstrong:2020woi,Albayrak:2020fyp,Adamo:2020qru,Alday:2021odx,Diwakar:2021juk,Drummond:2022dxd,Herderschee:2022ntr,Zhou:2021gnu,Sivaramakrishnan:2021srm,Alday:2022lkk,Cheung:2022pdk,Bissi:2022wuh,Li:2022tby,Lee:2022fgr,CarrilloGonzalez:2024sto,Beetar:2024ptv}. While a deep understanding of the color-kinematics duality is perhaps still missing, such as a yet-to-be-discovered kinematic Lie algebra~\cite{Monteiro:2011pc,Chen:2019ywi,Ben-Shahar:2021doh,Ben-Shahar:2021zww,Ben-Shahar:2022ixa,Ben-Shahar:2024dju}, a variety of approaches provide derivations of its tree-level consequences, ranging from string theory to scattering equations and positive geometry~\cite{BjerrumBohr:2009rd,Stieberger:2009hq,Cachazo:2012uq,Arkani-Hamed:2017mur,Mizera:2019blq,Britto:2021prf,Ahmadiniaz:2021fey,Ahmadiniaz:2021ayd}.

The double copy~\cite{Bern:2008qj,Bern:2010ue} provides a deep link between color-kinematics duality in gauge theories and the observed product structure of gravitational amplitudes. The Kawai-Lewellen-Tye (KLT) relations~\cite{Kawai:1985xq} gave the first tree-level realization of string-theory gravitational amplitudes as products of color-ordered gluon amplitudes (open strings). In the field-theory limit, one obtains amplitudes in general relativity (GR) from YM theory. The double copy, phrased in terms of cubic duality-satisfying gauge-theory graphs that have their color factors replaced by a second copy of kinematic numerators, vastly generalized the scope of such relations. Any gauge theory that admits local kinematic numerator identities, dual to the necessary Jacobi and commutation identities of the Lie algebra color factors, can be used in the double copy. It is guaranteed to give consistent diffeomorphism-invariant gravitational amplitudes~\cite{Chiodaroli:2017ngp}. 

A crucial aspect of the double copy is that it automatically generalizes to gravitational loop amplitudes~\cite{Bern:2010ue,Bern:2011rj, BoucherVeronneau:2011qv, Bern:2013uka,Bern:2014sna,Chiodaroli:2015wal,Johansson:2017bfl,Chiodaroli:2017ehv,Bern:2018jmv,Bern:2021ppb,He:2015wgf,Geyer:2015jch}, assuming that the color-kinematics duality holds for the underlying gauge-theory loop amplitudes. The double copy also applies to more general gravitational theories~\cite{Broedel:2012rc, Chiodaroli:2013upa,Johansson:2014zca,Chiodaroli:2014xia,Johansson:2015oia,Chiodaroli:2015rdg,Johansson:2017srf,Chiodaroli:2018dbu,Johansson:2018ues,Azevedo:2018dgo,Johansson:2019dnu,Bautista:2019evw,Plefka:2019wyg,Pavao:2022kog, Mazloumi:2022nvi}, classical solutions~\cite{Monteiro:2014cda,Luna:2015paa,Luna:2016hge,Bahjat-Abbas:2017htu,Carrillo-Gonzalez:2017iyj,Berman:2018hwd,CarrilloGonzalez:2019gof,Goldberger:2019xef,Huang:2019cja,Bahjat-Abbas:2020cyb,Easson:2020esh,Emond:2020lwi, Godazgar:2020zbv,Chacon:2021wbr,Chacon:2020fmr,Alfonsi:2020lub, Monteiro:2020plf, White:2020sfn, Elor:2020nqe,Pasarin:2020qoa, Adamo:2021dfg, Easson:2022zoh,Dempsey:2022sls,CarrilloGonzalez:2022ggn} and even some non-perturbative solutions~\cite{Cheung:2022mix,Armstrong-Williams:2022apo}. 
Higher-genus string-theory calculations also exhibit useful double-copy structures~\cite{DHoker:1989cxq,Geyer:2015jch,Geyer:2016wjx,He:2016mzd,He:2017spx,Geyer:2019hnn,Casali:2020knc,Geyer:2021oox,Stieberger:2022lss}.
The so-called classical double copy \cite{Monteiro:2014cda, Luna:2015paa} has led to new understanding between exact solutions in gravity and gauge theory, such as the Schwarzschild black hole and the Coulomb solution. 
Recent studies have applied the double copy to dynamical black holes, simplifying advanced GR calculations for black-hole scattering and gravitational-wave physics~\cite{Luna:2016due,Goldberger:2016iau,Luna:2017dtq,Shen:2018ebu,Plefka:2018dpa,Bern:2019nnu,Plefka:2019hmz,Bern:2019crd,Bern:2020buy,Almeida:2020mrg,Haddad:2020tvs,Bern:2021dqo,Bern:2021yeh,Bern:2022kto, Chiodaroli:2021eug,Shi:2021qsb,CarrilloGonzalez:2022mxx,Cangemi:2022abk,Bjerrum-Bohr:2022ows,Comberiati:2022cpm,Elkhidir:2023dco,Georgoudis:2023lgf,Georgoudis:2023ozp}.

While the double copy gives consistent diffeomorphism-invariant amplitudes, the resulting gravitational process may contain states and interactions that are undesired. The well-known example is the double copy of pure YM, which contains not only gravitons but also a scalar dilaton and pseudo-scalar axion (or $B^{\mu \nu}$ tensor in general dimension). Since the double copy is used for computing GR amplitudes, this is a potential problem. However, the dilaton and axion will automatically decouple in tree-level amplitudes with only external gravitons. They cannot be sourced by the gravitons; only pair-production of dilatons and axions are allowed, or equivalently, the states can form closed loops in loop-level amplitudes. 

The situation changes if we allow for massive matter in the double copy, which now sources the dilaton and potentially also the axion. In \rcites{Carrasco:2020ywq,Carrasco:2021bmu}, this problem was systematically addressed at tree and loop level for massive scalar matter, which sources only dilatons (and gravitons). By using an on-shell state projection to remove the dilatons, GR quantum amplitudes for massive spinless matter was explicitly computed up to the one-loop five-point level, from corresponding YM amplitudes. The results demonstrated that the combination of the double-copy, projection and unitarity methods provided a robust approach for obtaining non-trivial quantum amplitudes in GR, as well as in other theories with similar unwanted states. Later, the one-loop five-point amplitude of \rcite{Carrasco:2021bmu} served as input for calculating the next-to-leading order waveform for the classical scattering of two Schwarzschild black holes~\cite{Elkhidir:2023dco,Georgoudis:2023lgf,Georgoudis:2023ozp}. The use of the combined double-copy, projection and unitarity methods was also crucial for the third and fourth post-Minkowskian calculations of binary black-hole scattering~\cite{Bern:2019crd,Bern:2021dqo}, which became extra powerful when using the simplifications of the classical limit.
For the full quantum amplitudes the problem of unwanted states is more computationally demanding, and finding improved methods is desirable.

In this paper, we develop a double-copy approach that should automatically remove the unwanted dilatons, without relying on projectors, in the gravitational scattering of spinless massive matter.  To this end, we introduce an additional scalar \ghost{} field in the underlying YM gauge theory, such that its double copy mimics the unwanted contributions of the dilaton. By assigning ghost-like statistical signs in the double copy, the new contributions can be engineered to cancel out the dilation. The approach of canceling the dilaton using ghosts is similar to \rcite{Johansson:2014zca}, except that here we consider massive interactions, bosonic \ghost{} states, and work in general spacetime dimension. Similar bosonic \ghost{} contributions have appeared in~\rcite{Luna:2017dtq}.  
We use both top--down approach of dimensional compactification, and a bottom--up bootstrap approach for tree- and loop-level amplitudes, to constrain the gauge theory interactions. We verify that the resulting tree-level double-copy amplitudes, with up to six massive external scalars, are indeed those of GR, where \rcite{Carrasco:2021bmu} provides the GR results. We also compute the one-loop four-scalar amplitude using the same \ghost{} double copy, and after removing the axion bubble graph by hand, we get a gravitational amplitude that differs from GR by a minor remainder. 
The remainder is a contact-term bubble graph that vanish in the classical limit, and in the concluding section we briefly discuss several possible options for removing it.

The paper is structured as follows:  In \msecref{Section2}, we introduce the basic framework for understanding dilaton amplitudes:
Three-point interactions, that serve as building blocks for the higher-point bootstrap, are discussed in sections \ref{Section:Why dilatons} and \ref{Sec:Three-point} for gravity and gauge theory, respectively.  Generalities of tree amplitudes are found in \msecref{sec:GeneralitiesAmplitudes}, and a gauge-theory Lagrangian obtained from compactification is discussed in 
\msecref{Sec: Dimensional compactification}.
 In \msecref{Sec:Four-point}, we use the bootstrap method to work out four-point amplitudes with a variety of external states, and investigate the detailed numerator relations that enforces color-kinematics duality. In \msecref{Sec:six-point}, we bootstrap the massive six-scalar tree amplitude, and in \msecref{Sec:one-loop} we bootstrap the four-point one-loop amplitude.

\section{The root of the dilaton problem}\label{Section2}

In this section, we briefly review the need to remove dilatons and then proceed to introduce appropriate gauge-theory scalar fields through dimensional reduction. We infer the three-point amplitudes and a Lagrangian with some free parameters, which will later be fixed.

\subsection{The appearance of dilatons}\label{Section:Why dilatons}

It is well known that tree-level graviton amplitudes in general relativity can be constructed from the double copy of corresponding gluon amplitudes in Yang-Mills theory.  However, in principle, the double copy also contains massless states beyond those of gravitons -- namely a scalar dilaton and a pseudo-scalar axion (anti-symmetric tensor in general dimension). 

Consider the state decomposition of the three-point amplitude,
\begin{equation}
\begin{aligned}
\left(
    \begin{gathered}
        \includegraphics[scale=1.2]{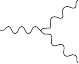}
    \end{gathered}
    \right)^2
    &=~
    \begin{gathered}
        \includegraphics[scale=1.2]{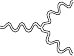}
    \end{gathered}    
~+~
     \begin{gathered}
        \includegraphics[scale=1.2]{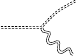}
    \end{gathered}
    ~+~
    \begin{gathered}
        \includegraphics[scale=1.2]{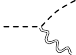}
    \end{gathered}
    ~+~
    \begin{gathered}
        \includegraphics[scale=1.2]{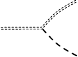}
    \end{gathered}
    ~+\text{perms}\,,
    \\
    (\text{YM})^2 \hspace{0.7cm} & 
    \hspace{0.7cm}
    \text{graviton} 
    \hspace{1.1cm}
    \text{axion}
    \hspace{1cm}
    \text{dilaton}
    \hspace{0.8cm}
    \text{axion/dilaton}
    \end{aligned}
\end{equation}
where ``perms'' refer to inequivalent permutations of the last three displayed graphs. Because 
the axion can only be pair-produced in a massless theory, it will automatically decouple from tree-level amplitudes with external gravitons. Likewise, in the absence of an axion, the dilaton can only be pair produced and thus it will also decouple from graviton tree amplitudes. 

Once massive matter is included in the double copy, the dilaton and axion can be sourced.  First, consider the double copy of a massive YM scalar at three points,
\begin{equation}
\begin{aligned}
\left(
    \begin{gathered}
        \includegraphics[scale=1.2]{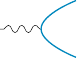}
    \end{gathered}
    \right)^2
    &~ =~~
    \begin{gathered}
        \includegraphics[scale=1.2]{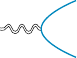}
    \end{gathered}    
~~+~~
    \begin{gathered}
        \includegraphics[scale=1.2]{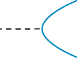}
    \end{gathered}
    ~,
    \\
    (\text{YM+scalar})^2 \hspace{0.0cm}  & 
    \hspace{0.85cm}
    \text{graviton} 
    \hspace{1.3cm}
    \text{dilaton}
    \end{aligned}
\end{equation}
where blue (or later red/green) solid lines from here on represent massive scalars. Since the double copy is a symmetric square, the massive scalar will source the dilaton, but not the axion. In contrast, if we consider a massive spinning particle, such as a fermion, and make the double copy asymmetric, both the axion and dilaton will be sourced,
\begin{equation}
\begin{aligned}
 \left(
    \begin{gathered}
        \includegraphics[scale=1.2]{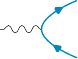}
    \end{gathered}
    \right)\times \left(
    \begin{gathered}
        \includegraphics[scale=1.2]{figures/mass_YM_3p.pdf}
    \end{gathered}
    \right) 
    &~ =~~
    \begin{gathered}
        \includegraphics[scale=1.2]{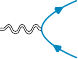}
    \end{gathered}    
~+~
     \begin{gathered}
        \includegraphics[scale=1.2]{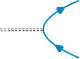}
    \end{gathered}
    ~+~
    \begin{gathered}
        \includegraphics[scale=1.2]{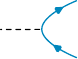}
    \end{gathered}
    ~.
    \\
  (\text{YM+fermion}) \times (\text{YM+scalar})  \hspace{0.0cm}  & 
    \hspace{0.75cm}
    \text{graviton} 
    \hspace{1.1cm}
    \text{axion}
    \hspace{1.3cm}
    \text{dilaton}
    \end{aligned}
\end{equation}
In this paper, we will only consider the former case, thus focusing on scalar sources and their couplings to dilatons. For any multiplicity, axions will not be sourced by massive scalars, thus we can ignore them at tree level.

While the above discussion was pictorial, the formal state decomposition is straightforward and can be applied directly to the outer product of two gluon polarization vectors $\varepsilon^\mu(k,q)$ and $\tilde \varepsilon^\mu(k,q)$,\footnote{Note that the sign in front of the projector $P^{\mu \nu}$ depends on the metric signature, we use mostly-minus metric $\eta_{\mu \nu}={\rm diag}(+,-,-,\ldots,-)$.}
\begin{equation} \label{GRstateDecomp}
\varepsilon^{\mu}\Tilde{\varepsilon}^{\nu} = \underbrace{
\frac{1}{2}\left(
\varepsilon^{\mu}\Tilde{\varepsilon}^{\nu}
+
\varepsilon^{\nu}\Tilde{\varepsilon}^{\mu}
+
\frac{2}{D_s -2} P^{\mu \nu}
    \right)
    }_{\text{graviton}}
    +
    \underbrace{
\frac{1}{2}\left(
\varepsilon^{\mu}\Tilde{\varepsilon}^{\nu}
-
\varepsilon^{\nu}\Tilde{\varepsilon}^{\mu}
    \right)
    }_{{\rm axion}/B^{\mu \nu}}
    -
    \underbrace{
\frac{1}{D_s -2} P^{\mu \nu}
    }_{\text{dilaton}}.
\end{equation}
where $P^{\mu \nu}=\eta^{\mu \nu}-2k^{(\mu} q^{\nu)}/k\cdot q$ is the transverse projector (or gluon state projector). The parameter $D_s$ denotes the spacetime dimension of the states, and in general it is taken to be independent of the spacetime dimension of the momenta $D=4-2\epsilon$. 
The contamination from dilatons and axions can be removed in order to obtain pure gravity amplitudes~\cite{Carrasco:2021bmu, Johansson:2014zca,Luna:2017dtq, Luna:2016hge}. While these states automatically decouple from graviton tree amplitudes, this does not happen at loop level and for tree amplitudes involving several massive external lines. The simplest example of the latter, is the four-point amplitude with two massive scalar lines,
\begin{equation}
    \begin{gathered}
        \includegraphics[scale=1.5]{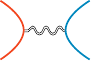}
    \end{gathered}.
\end{equation}
No projector acting on the external states can control the internal states propagating between the two massive lines. However, in this case one can still use the \textit{projective double copy} \cite{Carrasco:2021bmu} which analyses every factorization limit of the double copy amplitude using physical state projectors on internal lines, and then manually removes the unwanted dilaton/axion states (through the method of maximal cuts~\cite{Bern:2007ct}). The projective double copy is a powerful tool that can be used at any multiplicity both at tree and loop level and removes the need for gravitational amplitude ans\"{a}tze. However, as the resulting expressions are expanded squares (from the double-copies) they can get needlessly large and computations can be slow at high multiplicity and loop level. 

In this paper, we systematically study an alternative method for removing dilatons sourced by massive lines using a new double copy that targets the dilaton exchanges -- namely the \textit{\ghost{} double copy}. To this end, we introduce a massless scalar in the gauge theory that we call the \ghost, and as the name suggests, its double copy gives rise to a dilaton.\footnote{The double copy of the \ghost{} is more precisely a dilaton dressed with statistical signs that makes it a ghost.} The above massive scalar amplitude, with only graviton exchange, is then given in terms of a difference of two double copies,
\begin{equation}\label{Eq: Gravity prescription}
\begin{aligned}
    \begin{gathered}
        \includegraphics[scale=1.5]{figures/4pointtree_thick}
    \end{gathered}
    &= 
    \left(
    \begin{gathered}
        \includegraphics[scale=1.5]{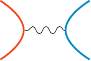}
    \end{gathered}
    \right)^2
    -
 \left(
    \begin{gathered}
        \includegraphics[scale=1.5]{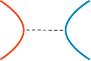}
    \end{gathered}
    \right)^2.
    \\
    \text{GR}
    \hspace{0.8cm}&
    \hspace{1.7cm}
 \text{YM}^2
    \hspace{2.2cm}
    \big(\text{\ghost}\big)^2
\end{aligned}
\end{equation}
As previously mentioned, we can ignore the axion since it is not sourced by massive scalars. The key advantage of this method is that each of the contributions are significantly simpler than the difference/sum of graphs. We will give the explicit expressions in later sections, after having introduced some general formalism.

\subsection{Generalities of tree-level amplitudes}\label{sec:GeneralitiesAmplitudes}
We now briefly discuss the notation for general tree amplitudes, to set the stage for later sections. For Yang-Mills amplitudes, we use calligraphic $\mathcal{A}$ for color-dressed amplitudes,
\begin{equation}
    \mathcal{A}_n^{\rm tree} = g^{n-2}\sum_{\Gamma \in \text{graphs}} \frac{C_\Gamma N_\Gamma}{\prod_{i\in \Gamma} (p_i^2-m_i^2)} = g^{n-2}\sum_{\sigma \in S_{n-2}} C(1,\sigma,n) A(1,\sigma,n)\, ,
\end{equation}
where $g$ is the coupling constant, $n$ is the multiplicity, $C_\Gamma$ and $N_\Gamma$ are the color factor and kinematic numerator of the graph $\Gamma$, respectively. The label $i$ runs over the the graph's propagators with momenta $p_i$ and mass $m_i$; the latter are zero for force carriers. For color-stripped partial amplitudes, we use $A(1,\sigma,n)$ where the ordering of labels corresponds to planar amplitudes, or equivalently for color factors in the adjoint representation, this corresponds to the Del~Duca-Dixon-Maltoni (DDM) half-ladder factors $C(1,\sigma,n)$. For general matter gauge-group representation, we still use planar partial amplitudes, and their color factors are now generalizations of the DDM factors, as given in \rcites{Johansson:2015oia,Melia:2015ika, Kalin:2017oqr, Ochirov:2019mtf}.  

For a general gravitational amplitude, we consider the double copy to be obtained from the Yang-Mills amplitude through the replacements $C_\Gamma \to \tilde N_\Gamma$, $g\to \kappa/2$,
\begin{equation}
    \mathcal{M}_n^{\rm tree} =  \Big(\frac{\kappa}{2}\Big)^{n-2} M(1,2,\ldots,n) =\Big(\frac{\kappa}{2}\Big)^{n-2} \sum_{\Gamma \in \text{graphs}} \frac{N_\Gamma \Tilde{N}_\Gamma}{\prod_{i\in \Gamma} (p_i^2-m_i^2)}\,, 
\end{equation}
where the numerators $\tilde N_\Gamma$ satisfy the same Lie-algebraic identities as the color factors $C_\Gamma$~\cite{Bern:2008qj,Bern:2010ue,Bern:2019prr}, and  $M(1,2,\ldots,n)$ refers to the coupling-stripped amplitude. 

For tree-level amplitudes that involve massive scalars, we define the gravitational amplitude that corresponds to general relativity (GR) as the following dilaton-subtracted formula:
\begin{equation}\label{Eq: Gravity amp prescription}
  M_{\rm GR}(1,2,\ldots,n)=\sum_\Gamma \frac{N_\Gamma \Tilde{N}_\Gamma}{\prod_{i\in \Gamma} (p_i^2-m_i^2)} +\sum_{l=1}^{\lfloor n/2 \rfloor}  \frac{ (-1)^{l}}{(D_s-2)^{l}} \sum_{\Gamma_l}   \frac{N_{\Gamma_l} \Tilde{N}_{\Gamma_l}}{\prod_{i\in \Gamma_l} (p_i^2-m_i^2)},
\end{equation}
where $l$ is the number of contiguous dilaton lines in the graph (the upper bound of the sum is at most $\lfloor n/2 \rfloor$, but it can be smaller if external gravitons are considered). The first term corresponds to the naive double copy of YM with massive scalars. The second term corresponds to the subtraction of graphs $\Gamma_l$ obtained from the double copy of YM with $l$ contiguous lines of \ghost{} contributions. All unwanted dilaton lines need to be subtracted out from the naive double copy; however, due to potential over-cancellations this means that graphs with even number of dilaton lines are added with a plus sign, which explains the $(-1)^l$ factor. The $(D_s-2)^{-1}$ factor comes from the normalization of the dilaton projector. Note that while each dilaton propagator has this factor, each quadratic-in-dilaton vertex also comes with the local factor $(D_s-2)$, thus each contiguous dilaton line has one more propagator than vertices and the resulting effective factor is $(D_s-2)^{-1}$.  

For this paper, we will rarely give the full GR amplitude, as these are generally given in \rcite{Carrasco:2021bmu}. Instead we will focus on the dilaton contributions that make up the difference between the naive double copy and the GR amplitude. This  difference amplitude is then
\begin{equation}\label{Eq: diference_amplitude}
  M(1,2,\ldots,n)=-\!\sum_{l=1}^{\lfloor n/2 \rfloor}  \frac{ (-1)^{l}}{(D_s-2)^{l}} \sum_{\Gamma_l}   \frac{N_{\Gamma_l} \Tilde{N}_{\Gamma_l}}{\prod_{i\in \Gamma_l} (p_i^2-m_i^2)},
\end{equation}
where we also allow ourself to generalize to the case that some of the external particles can be dilatons.  While this is not needed in \eqn{Eq: Gravity amp prescription}, the generalization to arbitrary states in \eqn{Eq: diference_amplitude} is convenient since it streamlines the analysis of factorization and unitarity properties. 

In some situations it is convenient to introduce the notation for a gravitational numerator $\mathcal{N}_\Gamma \equiv N_\Gamma \Tilde{N}_\Gamma$ for the graph $\Gamma$. In the case that $\Gamma$ is a pictorial representation of the graph,  it is also convenient to write the gauge theory numerator as $N(\Gamma)$. In contrast, if the graph topology is clear from the context, we sometimes only indicate the external leg labels in the numerator argument $N_i(1,2,\dots, n)$, and instead the index $i$ enumerates a distinct topology. Finally, for a pictorial graph $\Gamma$ that is used to express an amplitude, we take the graph to mean that it includes the numerator and denominator factor, $\Gamma \cong N_\Gamma/\prod_{i}(p_i^2-m_i^2)$, but not color factor.  Similarly, for the double copy of pictorial graphs with dilatons $\Gamma_l$, we use a slight abuse of notation for the ``square'',
\begin{equation} \label{squareAbuse}
\big(\Gamma_l\big)^2 \cong \frac{1}{(D_s-2)^l}\frac{N^2_{\Gamma_l}}{\prod_{i}(p_i^2-m_i^2)}\,,
\end{equation}
where the propagators and dilaton  normalization factors are included.  

The amplitudes in this paper are constrained using factorization limits and unitarity cuts, and we will use the same terminology for the two related notions. We refer to a maximal cut as a product of three-point amplitudes, summed over relevant intermediate states, 
\begin{equation}
\mathfrak{C} = \sum_{\text{states}} A_3 A_3' \cdots A_3''\,.
\end{equation}
For (next-to)${}^k$-maximal cuts we use the notation $\mathfrak{C}^{{\rm N}^k{\rm MC}}$ and the partial amplitudes now include factors at higher multiplicity, such that the cut has $k$ fewer on-shell propagators compared to the maximal cut. 

We will verify the double copy of our \ghost{} amplitudes by analyzing the difference of cuts, where GR amplitudes known from \rcite{Carrasco:2021bmu} serve as input. Specifically, we will refer to these as \textit{difference cuts}, and they are defined to be
\begin{equation}\label{eq: dilaton difference}
\Delta\mathfrak{C}\equiv
    \mathfrak{C}_{\text{YM}^2}-\mathfrak{C}_{\text{GR}},
\end{equation}
where $\mathfrak{C}_{\text{YM}^2}$ is the cut coming from the naive double copy of YM coupled to massive scalars, and $\mathfrak{C}_{\text{GR}}$ is the cut coming from GR. The YM${}^2$ results are obtained from \rcite{Carrasco:2020ywq}. The difference cut is non-zero whenever dilatons or axions contaminate the naive double copy. 

\subsection{Three-point amplitudes from dimensional compactification}\label{Sec:Three-point}

A convenient starting point for building gauge-theory amplitudes, needed for the double copy to dilaton amplitudes, is pure Yang-Mills (YM) theory in higher dimensions. In this setup, the matter particles obtain their mass from the first Kaluza-Klein (KK) mode in dimensional compactification (or reduction), and \ghost{} states emerge as extra-dimensional components of the zero-mode gluon. 

Consider the color-ordered three-point amplitude for YM theory in terms of polarization vectors $\varepsilon_i$ and momenta $k_i$,
\begin{equation}
A^{\text{YM}}(1,2,3) = (\varepsilon _1 \cdot \varepsilon _2)( k_2\cdot \varepsilon _3)+(\varepsilon _1\cdot \varepsilon _3)( k_1\cdot \varepsilon _2)+(\varepsilon _2\cdot \varepsilon _3)( k_3\cdot \varepsilon _1)\,.
\end{equation}
We now extract massive and massless scalar states by some simple operations. We can add a KK mass to legs 2 and 3 and get an amplitude,
\begin{equation}
A^{\rm YM_{KK}}(1,2,3) = A^{\text{YM}}(1,2,3)\Big|_{\substack{
k_2 \rightarrow k_2 + \mu \\
k_3 \rightarrow k_3 - \mu}}
\end{equation}
where $\mu$ is an extra-dimensional spacial vector that squares to the mass, $\mu^2=-m^2$. 
We will consider three needed interactions that we can extract from $A^{\text{YM}}$ and $A^{\rm YM_{KK}}$,  
\begin{equation}
\begin{gathered}
\includegraphics[scale=1.0]{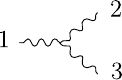}
\end{gathered}
\hspace{0.7cm}
\Rightarrow
\hspace{0.7cm}
\left\{
\hspace{0.2cm}
    \begin{gathered}
\includegraphics[scale=1.0]{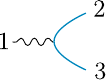}
\end{gathered}
\hspace{0.3cm}
,
\hspace{0.3cm}
 \begin{gathered}
\includegraphics[scale=1.0]{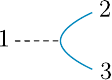}
\end{gathered}
\hspace{0.3cm}
,
\hspace{0.3cm}
\begin{gathered}
\includegraphics[scale=1.0]{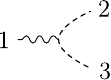}
\end{gathered}
\hspace{0.2cm}
\right\}
,
\end{equation}
where wavy lines represent gluons, dashed lines represent the \ghosts, and the solid blue lines represent massive scalars. The relevant expressions are extracted as follows:
\begin{align}\label{Eq: 3-point Higgs}
&\begin{gathered}
\includegraphics[scale=1.0]{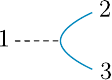}
\end{gathered}
\hspace{0.3cm}
:
\hspace{0.3cm}
    A(\g1,\b2,\b3) =\frac{\partial A^{\text{YM}_{\rm KK}}}{\partial \varepsilon_{23}}\Big|_{k_i=0} 
    \hspace{0.1cm}= 
   - \mu \cdot \varepsilon _1,
\\
\label{Eq: 3-point Higgs gluon}
&\begin{gathered}
\includegraphics[scale=1.0]{figures/gluon2dilaton_p.pdf}
\end{gathered}
\hspace{0.3cm}
:
\hspace{0.3cm}
A(1,\g2,\g3)
= 
  (\varepsilon_2 \cdot \varepsilon_3) \frac{\partial A^{\text{YM}}}{\partial \varepsilon_{23}}
  = (k_3\cdot \varepsilon_1) (\varepsilon_2 \cdot \varepsilon_3),
\hspace{0.5cm} \\
\label{Eq: 3-point mass gluon}
&\begin{gathered}
\includegraphics[scale=1.0]{figures/gluon_3p_p.pdf}
\end{gathered}
\hspace{0.3cm}
:
\hspace{0.3cm}
    A(1,\b2,\b3) =
    \hspace{0.1cm}
    \frac{\partial A_3^{\text{YM}_{\rm KK}}}{\partial \varepsilon_{23}}\Big|_{\mu=0}
   \hspace{0.2cm} =
   k_3\cdot \varepsilon _1 ,
\end{align}
where the derivative with respect to $\varepsilon_{23}=\varepsilon_2\cdot \varepsilon_3$ guarantees that the particle species is preserved along legs 2 and 3. For the \ghost{} we only need one extra-dimensional component; however, we keep using general polarization vectors in most formulae, since it gives a convenient bookkeeping device for later exercises.\footnote{For the purpose of constructing a \ghost{} state, we can set $\varepsilon_2\cdot \varepsilon_3=-1$ from the start, however we find it pedagogical to be slightly more general at intermediate steps. In any final expression, we will always assume the \ghost{} has only one component.} Note that we will not use different notation for regular- and extra-dimensional polarizations, instead the difference will be clear from the context and the drawn diagrams. 

The above three-point exercise can be repeated at higher multiplicity to get candidate amplitudes for Yang-Mills coupled to massive scalars and the \ghost{} state. However, we will proceed in two complementary directions. First, we will consider dimensional compactification at the Lagrangian level, to get a candidate for the full gauge theory. Afterwards, we will apply a bootstrap method to build higher-multiplicity amplitudes from the above three-point ones, and in the process also verify and constrain the results of the double copy by matching to GR.

\subsection{Dimensional compactification and truncation at Lagrangian level\label{Sec: Dimensional compactification}}
Here we work out a candidate gauge-theory Lagrangian from dimensional compactification. Starting from pure YM theory with gauge field $A_\mu$ we may formally decompose it as
\begin{align}
A_\mu \to  \{A_\mu,\varphi_i, \phi_a,\bar \phi^a \}\,,
\end{align}
where we think of the scalars as higher-dimensional gluons subject to a flat metric decomposed into three blocks $\eta_{\mu \nu}\to {\rm diag}(\eta_{\mu \nu}, -\delta_{ij},-\delta_a^b)$, and we assume that the complex scalars carry Kaluza-Klein mass vectors $\mu_{ia}$, through $\partial_i \phi_a = \mu_{ia}\phi_a$, $\partial_i \bar \phi_a = -\mu_{ia}\phi_a$, whereas $\varphi_i$ are massless \ghost{} scalars.  

Starting from the YM term ${\rm Tr} (F_{\mu \nu})^2$ and plugging in the above decomposition gives the following massless-sector Lagrangian:
\begin{align} \label{L_rootDil}
{\cal L}_\text{YM+\ghost}~ =~ &{\rm Tr} \Big(-\frac{1}{4}(F_{\mu \nu})^2+ \frac{1}{2}(D_\mu\varphi_i)^2  +\frac{g^2}{4}[\varphi_i,\varphi_j]^2 \Big)\,,
\end{align}
where $F_{\mu \nu} = \partial_\mu A_\nu-\partial_\nu A_\mu+ i g [A_\mu,A_\nu]$, $D_{\mu} \varphi_j  = \partial_\mu \varphi_j + i g [A_\mu,\varphi_j]$, and for simplicity we assume that $\varphi_i$ transforms in the adjoint of the gauge group.  In principle, we could allow for a free parameter in front of the quartic scalar term, but we will find that the above coupling is the appropriate one for the double copy.

For the complex matter, we will assume it transforms in a complex representation of the gauge group (e.g. obtained from breaking the gauge group and correlating the resulting representations with the above Lorentz decomposition). However, it is simpler to first work out the adjoint matter case, and then later swap the representation. From the decomposition of the YM term, with KK masses, we get the following Lagrangian for the adjoint matter case:
\begin{align} \label{LadjointMatter}
{\cal L}^\text{adjoint}_\text{matter}
= &   {\rm Tr} \Big( |D_\mu \phi_a|^2 
+\big(\mu_{i a} \phi_a +g [\varphi_i,\phi_a]\big) \big({-}\mu_{i a} \bar\phi^a +g [\varphi_i,\bar\phi^a]\big)
 \nn \\
&
\hspace{3cm}+ f_1 g^2 [\phi^a, \varphi_i] [\varphi_i,\bar\phi^a] - f_2 g^2 [\phi_a,\bar \phi^a]^2\Big),
\end{align}
where the covariant derivative is defined in the same way as below \eqn{L_rootDil}.
The first line of \eqn{LadjointMatter} comes from the dimensional compactification, and on the second line two free quartic-scalar couplings are added by hand, since these cannot be deduced by the preliminary three-point analysis of the previous section.

The factor $f_1$ will be determined in section~\ref{Sec:six-point} by comparing the double copy at six points to the corresponding GR amplitude. However, let us already now reveal the result of that analysis: we find that $f_1=\pm  \sqrt{D_s-1}$. For the $f_2$ factor, in principle one has the freedom to set it to any value since it will not alter classical physics when the scalar amplitudes are used to model scattering of Schwarzchild black holes; however, it is convenient to set $f_2=0$ to be consistent with the amplitudes already presented in \rcites{Carrasco:2021bmu,Carrasco:2020ywq}. This is also consistent with the case where we consider the $\phi_a$ to be real scalars, which then kills the commutator in the $f_2$ term. In practice, we will always align the charge of the massive scalars with the sign of the $\mu$ masses, thus if we keep track of the latter we can avoid drawing arrows on the scalar lines, and in many ways this makes the complex scalars equivalent to real scalars.

We may now promote the massive $\phi_a$ scalars to be transforming in the fundamental representation of the gauge group, and then the matter Lagrangian, with $f_2=0$, can be written as 
\begin{align} \label{LagrangianFundMatter}
{\cal L}^\text{fund.}_\text{matter}
= & \sum_{a=1}^{\# {\rm scalars}}  \Big( |D_\mu \phi_a|^2 
- \big(\mu_{i a} \bar\phi^a +g \bar\phi^a\varphi_i\big)\big(\mu_{i a} \phi_a +g \varphi_i\phi_a\big) + f_1 g^2 \bar\phi^a  \varphi_i \varphi_i \phi^a \Big)\, ,
\end{align}
where for clarity, we have spelled out the explicit sum over scalars of different masses. Diagrammatically, the massive and massless fields are given as,
\begin{equation}
\begin{aligned}
\text{Massive} 
: \hspace{0.8cm}
\phi_1
\hspace{0.2cm}
\begin{gathered}
\includegraphics[scale=1]{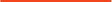}
\vspace{0.2cm}
\end{gathered} 
\hspace{1.2cm}
& 
\phi_2
\hspace{0.2cm}
\begin{gathered}
\includegraphics[scale=1]{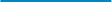}
\vspace{0.2cm}
\end{gathered} 
\\
\text{Massless} 
: \hspace{0.8cm}
\varphi_i 
\hspace{0.2cm}
\begin{gathered}
\includegraphics[scale=1]{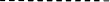}
\vspace{0.2cm}
\end{gathered} 
\hspace{1.2cm}
& 
A_{\mu}  
\hspace{0.2cm}
\begin{gathered}
\includegraphics[scale=1]{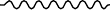}
\vspace{0.2cm}
\end{gathered}\, .
\end{aligned}
\end{equation}
Having completed the Lagrangian analysis, we now return to systematically studying the four- and higher-point amplitudes using the complementary approach of bootstrapping kinematic numerators.

\section{Four-point amplitudes from bootstrapping}\label{Sec:Four-point}

\begin{figure}
\centering
 \begin{subfigure}[c]{0.19\textwidth}
\includegraphics[scale=1.5]{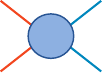}
\caption{$\mathcal{A}^{(\text{a})}$}
\end{subfigure}
\begin{subfigure}[c]{0.19\textwidth}
\includegraphics[scale=1.5]{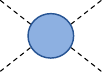}
\caption{$\mathcal{A}^{(\text{b})}$}
\end{subfigure}
\begin{subfigure}[c]{0.19\textwidth}
\includegraphics[scale=1.5]{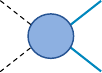}
\caption{$\mathcal{A}^{(\text{c})}$}
\end{subfigure}
\begin{subfigure}[c]{0.19\textwidth}
\includegraphics[scale=1.5]{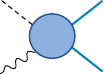}
\caption{$\mathcal{A}^{(\text{d})}$}
\end{subfigure}
\begin{subfigure}[c]{0.19\textwidth}
\includegraphics[scale=1.5]{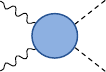}
\caption{$\mathcal{A}^{(\text{e})}$}
\end{subfigure}
\caption{The four-point amplitudes involving \ghosts{} and massive scalars, or gluons, or both.}
\label{Fig:All 4-point amps}
\end{figure}

We now go through the five distinct four-point amplitudes in \mfigref{Fig:All 4-point amps}, which need \ghost{} states in the double copy, and apply the bootstrapping method of \rcite{Carrasco:2020ywq}.  The first amplitude, $\mathcal{A}^{(\text{a})}$, is the simplest example of an amplitude involving a \ghost, and we will use it as an opportunity to elaborate on the bootstrapping method. The second amplitude, $\mathcal{A}^{(\text{b})}$, will demonstrate how \ghost{} states need the color-kinematic properties inherited from the gluons. The third amplitude, $\mathcal{A}^{(\text{c})}$, will introduce the need for a new modified BCJ relation for interactions with two \ghosts{} and two massive states. We use GR amplitudes obtained using projective double copy in \rcite{Carrasco:2021bmu} to verify that the \ghost{} double copy reproduces the so-called difference amplitudes \eqref{Eq: diference_amplitude}.

\subsection{Four massive scalars}\label{Sec: 4point same mass}

Consider the four-point amplitude involving two distinguishable massive scalars, interacting via a \ghost{} exchange. This amplitude is described by a single diagram,
\begin{equation}
A(\r1,\r2,\b3,\b4) = 
~
    \begin{gathered}
    \includegraphics[scale=1]{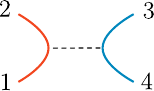}
    \end{gathered}
    ~
    = \frac{N(\r1,\r2,\b3,\b4)}{s_{12}}\,,
\end{equation}
where the Mandelstam variables are $s_{ij} = (k_i+k_j)^2$. The numerator function $N(\r1,\r2,\b3,\b4)$ contains all the non-trivial kinematic information, and we will now bootstrap it using an appropriate ansatz. It can be deduced from dimensional analysis that the numerator must be linear in dot products of the momenta $k_i\cdot k_j$ or masses $\mu_i\cdot \mu_j$, and there are no external polarizations. Since we expect the amplitude to vanish if either scalar becomes massless, the ansatz consists of a single term
\begin{equation} \label{num13}
\begin{aligned}
    N(\r1,\r2,\b3,\b4) =&~ c_1~(\mu_1 \cdot \mu_3) 
     ,
    \end{aligned}
\end{equation}
where $c_1$ is a coefficient to be fixed, and we recall that mass conservation implies that the mass-vectors are related as $\mu_2=-\mu_1$ and $\mu_4=-\mu_3$.

By unitarity, the amplitude must factorize into the product of the three-point amplitudes \eqref{Eq: 3-point Higgs} when the intermediate propagator goes on shell, $s_{12} = 0$. To sew together the amplitudes we sum over the appropriate states, which in this case is only the \ghost. Since we introduced the \ghost{} from dimensional reduction, we have to sum over the extra-dimensional states of a polarization vector, giving
\begin{equation}\label{Eq: ghost state projector}
   \sum_{\rm states} \varepsilon^{m}(k)     \varepsilon^{n}(-k) =  \eta^{mn}\,,
\end{equation}
where $m$, $n$ are now Euclidean indices, except we have to recall the slightly cumbersome choice of mostly-minus signature $\eta^{mn}={\rm diag}(-,-,-,\ldots,-)$. 
Using this metric and the three-point amplitudes \eqref{Eq: 3-point Higgs}, we find the cut of the factorization channel to be
\begin{equation}\label{Eq:4point 4mass cut}
\mathfrak{C}(\r1,\r2,\b3,\b4) =
    \begin{gathered}
   \includegraphics[scale=1.0]{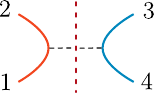}
    \end{gathered}
    = \sum_{\text{states}} A(\r1,\r2, \g{l})A(-\g{l}, \b3,\b4)
    = \mu _1\cdot \mu _3, 
\end{equation}
which fixes $c_1 =1$ in the ansatz.

\subsubsection*{Equal-mass amplitude}\label{Sec: 4point 4mass same mass}

It is interesting to discuss four-point scattering in the case of indistinguishable scalars. While this case in principle follows from the setting equal the masses of distinguishable scalars, and summing over channels, there are some noteworthy features, such as new relations between the diagrams.

Let us start by giving the numerators where the gluon propagates in the intermediate $s_{12}$ channel. The expressions are formally identical for the distinguishable and indistinguishable scalar cases,
\begin{equation}
N\left(
\begin{gathered}
    \includegraphics[scale=1.0]{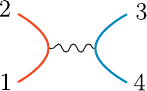}
\end{gathered}
    \right)
    = \frac{1}{2}k_2 \cdot (k_4 - k_3)
    =
    N\left(
\begin{gathered}
    \includegraphics[scale=1.0]{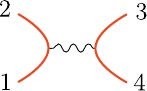}
\end{gathered}
    \right);
\end{equation}
however, we have to remember that the on shell conditions for the momenta are somewhat different in the two cases.

The indistinguishable case admits two more channels, given by permutations of the external legs, and the numerators obey the following kinematic relation: 
\begin{equation}\label{Eq:massive jacobi}
N\left(
\begin{gathered}
    \includegraphics[scale=1.0]{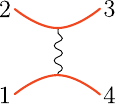}
    \end{gathered}
    \right)
    = 
N \left(
    \begin{gathered}
    \includegraphics[scale=1.0]{figures/4massglue_samemass.pdf}
\end{gathered}
\right)
-
N \left(
    \begin{gathered}
    \includegraphics[scale=1.0]{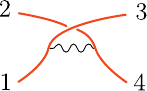}   
    \end{gathered}
    \right).
\end{equation}
If the matter transforms in the adjoint representation, we may view this as a kinematic Jacobi relation; however, since we prefer to use the fundamental representation, we will view it as kinematic bonus relation which has no analogue in terms of color factors.  

For the corresponding \ghost-mediated diagrams we have to be a bit more careful in the indistinguishable case. Setting equal the masses $\mu_1 = \mu_3=\mu$ in \eqn{num13}, one can find a similar kinematic relation between numerators in different channels.
However, since the mass stems from dimensional reduction, we need to keep track of the conserved $\mu$ flow. This can be done using complex scalars where the charge and $\mu$ direction are aligned (as in section \ref{Sec: Dimensional compactification}), or using real scalars with $\mu$ flow indicated similarly to momentum (which we use from here on). This makes it possible find the following kinematic two-term identity:
\begin{equation} \label{Mu2termID}
N \left(
     \begin{gathered}
        \includegraphics[scale=1.0]{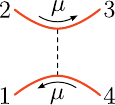}
    \end{gathered}
\right)
    =
N\left(
    \begin{gathered}
        \includegraphics[scale=1.0]{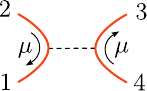}
    \end{gathered}
\right)
= \mu^2,
\end{equation}
With the above arrow assignments there is no way to write down the $s_{13}$-channel diagram; however, for other arrow assignments it does of course exist. For any choice of arrow assignments, the kinematic two-term identity will exist, and we will view it as a bonus relation, on similar footing as \eqn{Eq:massive jacobi}.

\subsubsection*{Gravity comparison}
We now verify that the double copy of the numerator in \eqn{Eq:4point 4mass cut} reproduces the correct difference cut, as defined in \eqn{eq: dilaton difference}. Thus we compute the only difference cut (or factorization limit) as
\begin{equation}
\begin{aligned}
\Delta \mathfrak{C}
&\equiv 
\hspace{0.15cm}
\Big[\mathfrak{C}_{\text{YM}}(\r1,\r2,\b3,\b4) \Big]^2 
\hspace{0.2cm}
- 
\hspace{0.1cm}
\mathfrak{C}_{\text{GR}}(\r1,\r2,\b3,\b4)
\\
&=
\left(
\begin{gathered}
    \includegraphics[scale=1.0]{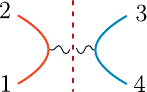}
\end{gathered}\right)^2
- 
\begin{gathered}
    \includegraphics[scale=1.0]{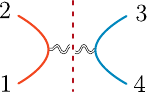}
\end{gathered}
\\
&=
\left(\frac{1}{2}k_2 \cdot  k_4\right)^2
-
\left(
\frac{1}{4}(k_2 \cdot  k_4)^2 - \frac{1}{D_s-2}m_1^2m_3^2
\right)= \frac{1}{D_s-2}\mu_1^2\mu_3^2\,,
\end{aligned}
\end{equation}
where the GR result comes from \rcite{Carrasco:2021bmu}, and we used $\mu^2_i=-m^2_i$. To be clear, the thin wavy line represents a gluon, and the wavy double  line represents a graviton here. Finally, restricting to one-dimensional mass vectors gives the identity $\mu_1^2\mu_3^2=(\mu_1\cdot\mu_3)^2$, and the difference cut matches the double copy \eqref{Eq: diference_amplitude} of the \ghost{} numerator \eqref{Eq:4point 4mass cut}.

\subsubsection*{Final expressions}
For the reader's convenience, we here summarize the obtained four-scalar amplitudes with \ghost{} and dilaton exchanges, they are
\begin{align}
    A(\r1,\r2,\b3,\b4) &= \frac{\mu_1 \cdot \mu_3}{s_{12}}\,,
\\
M(\r1,\r2,\b3,\b4) &= \frac{1}{D_s-2}\frac{\mu_1^2 \mu_3^2}{s_{12}}\,,
\end{align}
and, as demonstrated, they are related by the double copy.

\subsection{Two masses, two \ghosts{} and a new contact term}\label{sec:two mass two ghost}

A particularly interesting case is the two-\ghost{} two-massive amplitude, color coded as $A(\g1,\g2,\b3,\b4)$, which can be expressed in terms of the two graphs
\begin{equation}
\begin{aligned} 
\begin{gathered}
    \includegraphics[scale=1.0]{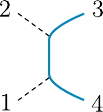}
\end{gathered}
\hspace{0.7cm}
&,
\hspace{0.7cm}
\begin{gathered}
    \includegraphics[scale=1.0]{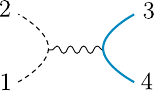}
\end{gathered}
~,
\\
N_1(\g1,\g2,\b3,\b4) 
\hspace{0.5cm}
&
\hspace{1cm}
N_2(\g1,\g2,\b3,\b4)
\end{aligned}
\end{equation}
where $N_i(\g1,\g2,\b3,\b4)$ are the corresponding kinematic numerator functions. Having more than one channel, we can now try to relate the numerators using a novel three-term identity, which we view as a bonus relation since it is not needed for enforcing diffeomorphism invariance of the double copy. The bonus numerator relation has the somewhat unusual form
\begin{equation}\label{Eq: Jacobi d2m2}
\begin{aligned}
\fD \times N \left(
    \begin{gathered}
    \includegraphics[scale=1.0]{figures/2mass2dilaton_s_p.pdf}
\end{gathered}
\right)
&=
N \left(
\begin{gathered}
~
    \includegraphics[scale=1.0]{figures/2mass2dilaton.pdf}
    ~
\end{gathered}
\right)
-
N
\left(
\begin{gathered}
~
    \includegraphics[scale=1.0]{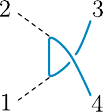}
    ~
\end{gathered} 
\right)
,
\end{aligned}
\end{equation}
where a multiplicative factor $\fD$ appears in front of the graph with the single contiguous \ghost{} line. The factor will be determined by comparing the double-copy of six-point amplitudes to known GR results (see \msecref{Sec:six-point}), and we will slightly spoil that analysis by giving the result: 
\begin{equation}\label{Eq:fD}
\fD = -1 \pm\sqrt{D_s-1}= -1+ f_1,    
\end{equation}
where $f_1$ is the factor that appeared in the Lagrangian \eqref{LagrangianFundMatter}. In terms of standard notation, the new bonus relation is
\begin{equation}\label{Eq: Jacobi d2m2 numerators}
    \fD \times N_2(\g1,\g2,\b3,\b4) =
N_1(\g1,\g2,\b3,\b4) 
-
N_1(\g1,\g2,\b4,\b3)\, .
\end{equation}
The two sign choices of $\fD$ will be reflected in our use of an asymmetric double copy, where the $N_i$ numerators will use $\fDp$ and the $\tilde N_i$ numerators use $\fDm$, and in the product the square root will become rationalized, e.g. $\fDp\fDm =-(D_s-2)$. This relative factor between the numerators in \eqn{Eq: Jacobi d2m2} is actually natural, since in the double copy we need to account for dilaton normalization factors $(D_s-2)^{-1}$ that differ for graphs with one or two contiguous \ghost{} lines.

From \eqn{Eq: Jacobi d2m2 numerators} it is clear that the numerator $N_2(\g1,\g2,\b3,\b4)$ can be fully expressed in terms of $N_1(\g1,\g2,\b3,\b4)$. We give this numerator an ansatz that is constrained to satisfy the graph automorphism $N_1(\g2,\g1,\b4,\b3)= N_1(\g1,\g2,\b3,\b4)$,
\begin{equation}
\begin{split}
  N_1(\g1,\g2,\b3,\b4) =& \hspace{0.2cm}
\varepsilon_1\cdot \varepsilon_2 \left(c_1 k_2\cdot k_4+c_2 k_3\cdot k_4+c_3 \mu ^2 \right)
+c_4(\varepsilon_1\cdot \mu)(\varepsilon_2\cdot \mu).
\end{split}
\end{equation}
Next, the remaining free coefficients $c_i$ will be fully fixed by the two factorization channels of the amplitude.

\subsubsection*{The factorizations}
The gluon factorization channel is given by 
\begin{equation}
\begin{aligned}
     \begin{gathered}
    \includegraphics[scale=1.0]{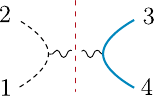}
\end{gathered}
&= \sum_{\text{states}}
A(\g1,\g2, l)A(-l,\b3,\b4)= \frac{1}{2}(\varepsilon_1\cdot \varepsilon_2 ) ~t_{24},
\end{aligned}
\end{equation}
where $t_{ij} =  2 (k_i \cdot k_j)$.  The massive factorization channel is given by 
\begin{equation}
\begin{aligned}
    \begin{gathered}
    \includegraphics[scale=1.0]{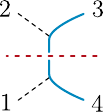}
\end{gathered}
=& \sum_{\text{states}} A(\b4,\g1, \b{l}) A(-\b{l},\g2,\b3)
= ~(\mu \cdot \varepsilon_1)( \mu \cdot \varepsilon_2)\,.
\end{aligned}
\end{equation}
where $\mu=\mu_3=-\mu_4$. We compare these expressions to the ansatz and use the equalities to fix the numerators. 

\subsubsection*{Final expressions}
Using the information of the factorization channels, the final numerator functions are given by
\begin{equation}\label{Eq: 2d2m final numerator}
  N
  \left(
   \begin{gathered}
    \includegraphics[scale=1.0]{figures/2mass2dilaton.pdf}
\end{gathered}
  \right)
  =
  N_1(\g1,\g2,\b3,\b4)
 =(\varepsilon_1\cdot \mu)(  \varepsilon_2\cdot \mu)
 -\frac{\fD}{4} (\varepsilon_1\cdot \varepsilon_2)t_{14} ,
\end{equation}
and
\begin{equation} \label{Eq: 2d2m second final numerator}
\begin{aligned}
N\left(
\begin{gathered}
    \includegraphics[scale=1.0]{figures/2mass2dilaton_s_p.pdf}
\end{gathered}
\right)= 
N_2(\g1,\g2,\b3,\b4) = (\varepsilon_1\cdot \varepsilon_2)\frac{t_{24}-t_{14}}{4},
    \end{aligned}
\end{equation}
where the latter is given by the bonus relation \eqref{Eq: Jacobi d2m2 numerators}.

To be explicit, the color-ordered gauge-theory amplitude is given by
\begin{equation}
\begin{aligned}
A(\g1,\g2,\b3,\b4) =& \hspace{0.5cm}
 \begin{gathered}
    \includegraphics[scale=1.0]{figures/2mass2dilaton.pdf}
\end{gathered}
\hspace{0.5cm}
+
\hspace{0.5cm}
 \begin{gathered}
    \includegraphics[scale=1.0]{figures/2mass2dilaton_s_p.pdf}
\end{gathered}
\\
=&~
\frac{(\varepsilon_1\cdot \mu )( \varepsilon_2\cdot \mu )}{t_{14}}+\frac{\varepsilon_1\cdot \varepsilon_2}{2} \left(1 \pm\frac{1}{2}\sqrt{D_s-1}+\frac{ t_{24}}{s_{12}}\right) .
\end{aligned}
\end{equation} 
Finally, we can assemble the corresponding gravitational double copy as\footnote{Since the amplitude \eqref{TowDilatonAmp} has external dilatons, one can change the overall normalization by rescaling the dilaton field by suitable $(D_s-2)^\alpha$ factors to obtain canonically normalized kinetic terms and amplitudes.}
\begin{equation} \label{TowDilatonAmp}
    \begin{aligned}
    M(\g1,\g2,\b3,\b4) =& 
   - \left|
    ~
    \begin{gathered}
    \includegraphics[scale=1.0]{figures/2mass2dilaton.pdf}
\end{gathered}
~
\right|^2
-
\left|
\begin{gathered}
    \includegraphics[scale=1.0]{figures/2mass2dilaton_u.pdf}
\end{gathered}
\right|^2
+
\left(
 \begin{gathered}
    \includegraphics[scale=1.0]{figures/2mass2dilaton_s_p.pdf}
\end{gathered}
\right)^2
\\
=& ~
\frac{1}{(D_s-2)^2}\left(\frac{(\mu \cdot \varepsilon_1)^2(\mu \cdot \varepsilon_2)^2 s_{12}}{t_{24} t_{14}}-(\varepsilon_1\cdot \varepsilon_2 )(\varepsilon_1\cdot \mu  )(\varepsilon_2\cdot \mu )\right)\\
&\hspace{5cm}
-\frac{1}{ D_s-2}\frac{t_{24} t_{14} \left(\varepsilon_1\cdot \varepsilon_2\right){}^2}{4 s_{12}} \\
=&~ \frac{1}{(D_s - 2)^2} \Big(\frac{m^4 s_{12}}{t_{14} t_{24}} + m^2\Big)- \frac{1}{D_s - 2}\frac{t_{14} t_{24}}{4s_{12}}\,,
    \end{aligned}
\end{equation}
where on the last line we converted to the relevant case of dilaton states with only one extra-dimensional component ($\varepsilon_i\cdot \varepsilon_j= -1$, $\varepsilon_i\cdot \mu= -m$). Note that, as explained in \eqn{squareAbuse}, we use the abbreviated pictorial double-copy notation for the ``square'' of a graph $\Gamma_l$, 
\begin{equation}\label{Eq: 2* definition}
    \big(\Gamma_l\big)^2 \equiv \frac{1}{(D_s-2)^{l}}  \frac{N_{\Gamma_l}^2}{\prod_i (p_i^2-m_i^2)},~~~~~~~
    \big|\Gamma_l \big|^2 \equiv \frac{1}{(D_s-2)^{l}}
    \frac{N_{\Gamma_l} \tilde N_{\Gamma_l} }{\prod_i (p_i^2-m_i^2)},
\end{equation}
where $N_{\Gamma_l}$ depends on $\fDp$, $\tilde N_{\Gamma_l}$ depends on $\fDm$, and $l$ is the number of contiguous dilaton lines in the graph.

\subsection{Four \ghost{} states: inheriting color-kinematics duality}

To probe the color-kinematic properties of the \ghost\, amplitudes we take a close look at the four-\ghost\, amplitude, and investigate the features inherited from the pure YM amplitude. The amplitude $A(\g1,\g2,\g3,\g4)$ depends on permutations of a single graph
\begin{equation}
\begin{gathered}
\begin{aligned}
    & \begin{gathered}
    \includegraphics[scale=1.0]{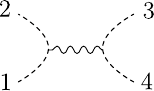}
    \end{gathered}
    \\
    & \hspace{0.3cm} N(\g1,\g2,\g3,\g4)
\end{aligned}
\end{gathered}
\hspace{0.5cm},
\end{equation}
where $N(\g1,\g2,\g3,\g4)$ is the unknown kinematic numerator function.

The ansatz for the  numerator function is similar to the pure YM gluon numerator,
\begin{align}
N(\g1,\g2,\g3,\g4)
     =& \, k_2\cdot k_4 \left(c_1 (\varepsilon _1\cdot \varepsilon _2)( \varepsilon _3\cdot \varepsilon _4)+c_2 (\varepsilon _1\cdot \varepsilon _3)( \varepsilon _2\cdot \varepsilon _4)+c_3 (\varepsilon _1\cdot \varepsilon _4)( \varepsilon _2\cdot \varepsilon _3)\right)\nn\\
     &+k_3\cdot k_4 \left(c_4 (\varepsilon _1\cdot \varepsilon _2)( \varepsilon _3\cdot \varepsilon _4)+c_5 (\varepsilon _1\cdot \varepsilon _3)( \varepsilon _2\cdot \varepsilon _4)+c_6 (\varepsilon _1\cdot \varepsilon _4)( \varepsilon _2\cdot \varepsilon _3)\right),
\end{align}
except we recall that the \ghost{}  polarizations should be thought of as extra-dimensional states, unlike the gluon ones.
Half of the free parameters are fixed by imposing the automorphism identities $N(\g1,\g2,\g3,\g4)=-N(\g1,\g2,\g4,\g3)=N(\g2,\g1,\g4,\g3)$. Next, we constrain the numerator from the gluon factorization channel,
\begin{equation}
    \begin{gathered}
    \includegraphics[scale=1.0]{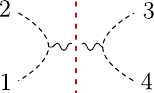}
    \end{gathered}
    = 
    \sum_{\text{states}}A(\g1, \g2, l)A(-l, \g3, \g4)
    = \frac{1}{2}s_{24}(\varepsilon_1\cdot \varepsilon_2 )(\varepsilon_3\cdot \varepsilon_4)\,.
\end{equation}
After imposing this factorization, we are left with a single free parameter $c_{\times}$ which controls a contact term,
\begin{equation}
\begin{aligned}
N(\g1,\g2,\g3,\g4)
=&   \frac{1}{4} (\varepsilon_1\cdot \varepsilon_2)( \varepsilon_3\cdot \varepsilon_4 )\left(s_{24}-s_{14}\right)\\
    &+ c_{\times} s_{12} \big[(\varepsilon_1\cdot \varepsilon_3)( \varepsilon_2\cdot \varepsilon_4)-(\varepsilon_1\cdot \varepsilon_4 )(\varepsilon_2\cdot \varepsilon_3)\big].
    \end{aligned}
\end{equation}
The free parameter can be fixed by imposing a kinematic Jacobi relation for the numerator,
\begin{equation}\label{Eq: 4 dilaton jacobi}
N
\left(
     \begin{gathered}
    \includegraphics[scale=1.0]{figures/4dilaton_p.pdf}
    \end{gathered}
    \right)
    = 
N
\left(
     \begin{gathered}
    \includegraphics[scale=1.0]{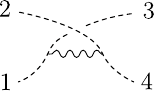}
    \end{gathered}
    \right)
+
N
\left(
     \begin{gathered}
    \includegraphics[scale=1.0]{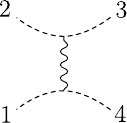}
    \end{gathered}
    \right)
    ,
\end{equation}
which determines $c_{\times} = \frac{1}{4}$. Note that this kinematic Jacobi identity is inherited from the dimensional reduction we use; however, in principle the identity is not needed for obtaining diffeomorphism invariant double-copy amplitudes.

\subsubsection*{Final expressions}

The fully fixed numerator function $N(\g1,\g2,\g3,\g4)$ is
\begin{equation}
\begin{aligned}
N \left(
\begin{gathered}
    \includegraphics[scale=1.0]{figures/4dilaton_p.pdf}
\end{gathered}
\right)
=   
\frac{1}{4} \Big((s_{24}-s_{14}) \varepsilon_{12} \varepsilon_{34} + s_{12}\left(\varepsilon_{13} \varepsilon_{24} - \varepsilon_{14} \varepsilon_{23} \right)  \Big)\,,
    \end{aligned}
\end{equation}
where we recall that $\varepsilon_{ij}=\varepsilon_i\cdot \varepsilon_j$.
The color-ordered gauge-theory amplitude is then
\begin{equation}
\begin{aligned}
    A(\g1,\g2,\g3,\g4) =&
    \hspace{0.6cm}
     \begin{gathered}
    \includegraphics[scale=1.0]{figures/4dilaton_p.pdf}
    \end{gathered}
    \hspace{0.5cm}
    +
    \hspace{0.5cm}
     \begin{gathered}
    \includegraphics[scale=1.0]{figures/4dilaton_s_p.pdf}
    \end{gathered}\\
    =&~
\frac{1}{4} \left(\varepsilon_{13} \varepsilon_{24}-\varepsilon_{14} \varepsilon_{23}+\frac{\varepsilon_{12} \varepsilon_{34} \left(s_{24}-s_{14}\right)}{s_{12}}\right)
\\
&\hspace{2cm}+\frac{1}{4} \left(\varepsilon_{13} \varepsilon_{24}-\varepsilon_{12} \varepsilon_{34}+\frac{\varepsilon_{14} \varepsilon_{23} \left(s_{24}-s_{12}\right)}{s_{14}}\right),
    \end{aligned}
\end{equation}
and the gravitational four-dilaton amplitude is given by the double copy
\begin{equation}
    \begin{aligned}
    M(\g1,\g2,\g3,\g4) =&
    \hspace{0.2cm}
    \left(
     \begin{gathered}
    \includegraphics[scale=1.0]{figures/4dilaton_p.pdf}
    \end{gathered}
    \right)^{\hspace{-0.2cm}2}
    \hspace{0.1cm}
    +
    \hspace{0.2cm}
    \left(    
    \begin{gathered}
    \includegraphics[scale=1.0]{figures/4dilaton_u_p.pdf}
    \end{gathered}
    \right)^{\hspace{-0.2cm}2}
    \hspace{0.1cm}
    +
    \hspace{0.2cm}
    \left(
     \begin{gathered}
    \includegraphics[scale=1.0]{figures/4dilaton_s_p.pdf}
    \end{gathered}
    \right)^{\hspace{-0.2cm}2}
    \\
    =& ~
   \frac{-1}{(D_s-2)^2}   \frac{\left(\left(\varepsilon_{14} \varepsilon_{23}-\varepsilon_{13} \varepsilon_{24}\right) s_{12}+\varepsilon_{12} \varepsilon_{34} \left(s_{14}-s_{24}\right)\right){}^2}{16 s_{12}}
    + \text{perms}(1,2,3)\\
        =& ~
  - \frac{
     \left(s_{14}-s_{24}\right){}^2}{16 (D_s-2)^2 s_{12}} 
     + \text{perms}(1,2,3)
 = \frac{1}{(D_s-2)^2}\frac{\left(s_{12}^2+s_{24}^2 +s_{14}^2\right)^2}{16  s_{12}s_{24} s_{14}}.
    \end{aligned}
\end{equation}
The last line comes from using that $\varepsilon_{ij}=-1$, which is the relevant case for a single dilaton state. As before, the overall normalization follows our conventions for the \ghost{} double copy, and it differs from canonical normalization by $(D_s-2)$ factors.

\subsection{Summary: four-point amplitudes and observations}

For the reader's convenience, the distinct graph topologies and numerators that enter the discussed four-point amplitudes are summarized in \mtabref{tab:four_point}. While not discussed in any detail, in \mtabref{tab:five_point} we give the graph topologies and numerators of the \ghost-mediated five-point amplitude involving massive scalars and one external gauge boson. The double copy of this five-point amplitude agrees with the results from \rcite{Luna:2017dtq}. The following novel lessons and observations from the determined \ghosts{} interactions are worth stressing: 
\begin{itemize}

\item In addition to the known kinematic Jacobi relations between numerators, there exist bonus relations which we take advantage of in our bootstrap. In particular, we find a novel type of numerator relation that involves a multiplicative factor that compensates for the normalization difference of one versus two contiguous \ghost{} lines,
\begin{equation}
\fD \times
N\left(
    \begin{gathered}
    \includegraphics[scale=1.0]{figures/2mass2dilaton_s_p}
\end{gathered}
\right)
=
N\left(
\begin{gathered}
    \includegraphics[scale=1.0]{figures/2mass2dilaton}
\end{gathered}
\right)
-
N \left(
\begin{gathered}
    \includegraphics[scale=1.0]{figures/2mass2dilaton_u}
\end{gathered}
\right)
,
\end{equation}
where $\fD = -1\pm\sqrt{D_s-1}$. Other more conventional bonus relations are \eqnss{Eq: 4 dilaton jacobi}{Eq:massive jacobi}{Mu2termID}, which are inherited from YM via the dimensional reduction approach we use. Regarding standard kinematic Jacobi relations for four-point numerators with external gluons, see \rcite{Carrasco:2020ywq} for details.

\item A \ghost{} line sourced by two scalars carrying mass vectors $\mu_i,\mu_j$, gives rise to interactions proportional to $\mu_i \cdot \mu_j$. In the double-copy amplitude one converts to one-dimensional dot products to obtain proper masses, e.g.  $(\mu_i \cdot \mu_j)^2= \mu_i^2  \mu_j^2= m_i^2  m_j^2$.

\item Double-copy diagrams are dressed with normalization factors $(-1)^{l-1}(D_s-2)^{-l}$, where $l$ are the number of contiguous dilaton lines that are sourced by massive lines. (For closed dilaton loops, the normalization factor is unity.)

\end{itemize}

\begin{table}[]
    \centering
    \begin{tabular}{
    M{0.1\linewidth}
    |M{0.25\linewidth}
    |M{0.5\linewidth}}
    \text{Ampl.} &\text{Topologies}  & \text{Numerator}\\
    \hline 
    $\mathcal{A}^{(\text{a})}$ 
    &     
    \vspace{-0.4cm}
     \begin{gather*}
     \begin{gathered}
     \includegraphics[scale=1.0]{figures/4massdil.pdf}
 \end{gathered} 
 \end{gather*}
 \vspace{-0.4cm}
&
\begin{gather*}
    \mu _1\cdot \mu _3
\end{gather*}
\\
\hline
$\mathcal{A}^{(\text{b})}$ 
&
\vspace{-0.4cm}
\begin{gather*}
\begin{gathered}
    \includegraphics[scale=1.0]{figures/4dilaton_p.pdf}
\end{gathered}
\end{gather*}
\vspace{-0.4cm}
&
\begin{gather*}
        \frac{1}{4} \Big[\left( s_{24}-s_{14}\right) \varepsilon_{12} \varepsilon_{34}
    +s_{12}(\varepsilon_{13} \varepsilon_{24}-\varepsilon_{14} \varepsilon_{23})  \Big]
    \end{gather*}
\\
\hline
$\mathcal{A}^{(\text{c})}$ 
&
\vspace{-0.4cm}
\begin{gather*}
    \begin{gathered}
   \includegraphics[scale=1.0]{figures/2mass2dilaton.pdf}    
\end{gathered}
\\
 \begin{gathered}
   \includegraphics[scale=1.0]{figures/2mass2dilaton_s_p.pdf} 
\end{gathered}
\end{gather*}
\vspace{-0.4cm}
&
\begin{gather*}
(\varepsilon_1\cdot \mu)(  \varepsilon_2\cdot \mu)
-\frac{1}{4} \fD \, t_{14}\varepsilon_{12}  
\\
\\
\\
\frac{1}{4} \left( t_{24}-t_{14} \right) \varepsilon_{12}
\end{gather*}
\\
\hline
$\mathcal{A}^{(\text{d})}$ 
&
\vspace{-0.4cm}
\begin{gather*}
\begin{gathered}
       \includegraphics[scale=1.0]{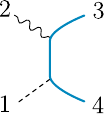}
\end{gathered}
\\
\begin{gathered}
      \includegraphics[scale=1.0]{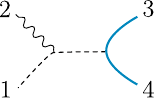}
\end{gathered}
\end{gather*}
\vspace{-0.4cm}
&
\begin{gather*}
( \varepsilon_1\cdot \mu )(\varepsilon_2 \cdot k_3)
\\
\\
\\
( \varepsilon_1\cdot \mu )(\varepsilon_2\cdot k_3)-( \varepsilon_1\cdot\mu)( \varepsilon_2 \cdot k_4)
\end{gather*}
\\
\hline
$\mathcal{A}^{(\text{e})}$ 
&
\vspace{-0.4cm}
\begin{gather*}
    \begin{gathered}
    \includegraphics[scale=1.0]{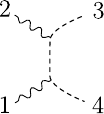}
\end{gathered}
\\
\begin{gathered}
    \includegraphics[scale=1.0]{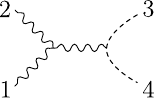}
\end{gathered}
\end{gather*}
\vspace{-0.4cm}
&
\begin{gather*}
    - \varepsilon_{34}~ \Big[ (\varepsilon_1 \cdot k_4)(\varepsilon_2 \cdot k_3) + s_{14} \varepsilon_{12}  \Big]
    \\
    \\
    \\
     \varepsilon_{12} ~\Big[(\varepsilon_3 \cdot k_1) (\varepsilon_4 \cdot k_2)-  ( \varepsilon_3 \cdot k_2)(\varepsilon_4 \cdot k_1)
    \\
    \hspace{2.5cm}
    +\frac{1}{4}\left(s_{24}-s_{23}\right) \varepsilon_{34}\Big]
\end{gather*}
\\
\end{tabular}
    \caption{A summary of the four-point amplitudes involving \ghosts, given in terms of the contributing graph topologies and kinematic numerators. The $\mu_i$ are mass vectors satisfying mass conservation and $\mu_i^2=-m_i^2$; they are assumed to be one-dimensional after evaluating the double copy.}
    \label{tab:four_point}
\end{table}

\begin{table}[]
    \centering
    \begin{tabular}{
    M{0.1\linewidth}
    |M{0.25\linewidth}
    |M{0.5\linewidth}}
    \text{Ampl.} &\text{Topologies}  & \text{Numerator}
\\
\hline
$\mathcal{A}_5$ 
&
\vspace{-0.4cm}
\begin{gather*}
    \begin{gathered}
    \includegraphics[scale=1.0]{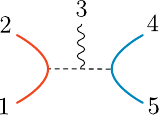}
\end{gathered}
\\
\begin{gathered}
    \includegraphics[scale=1.0]{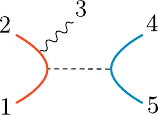}
\end{gathered}
\end{gather*}
\vspace{-0.4cm}
&
\begin{gather*}
    (\mu _1\cdot \mu _4) \left(k_1+k_2\right)\cdot \varepsilon_3
    \\
    \\
    \\
    (\mu _1\cdot \mu _4) \left(k_2\cdot \varepsilon_3\right)
\end{gather*}
\\
\end{tabular}
    \caption{A five-point amplitude with \ghost{} exchange, given in terms of the contributing graph topologies and kinematic numerators. }
    \label{tab:five_point}
\end{table}

\section{The six-point amplitude}\label{Sec:six-point}

Having completed the four-point analysis, we will jump directly to the six-point amplitude with only massive external scalars. This amplitude serves as the verification step for the lower-point interactions that we already discussed, since the external states are now all physical and can be compared to the corresponding GR amplitude. However, in practice, we both verify and constrain the lower-point interactions using the six-point amplitude, thus imposing that the \ghost{} double copy matches the difference cuts \eqref{eq: dilaton difference}, where the GR results come from \rcite{Carrasco:2021bmu}. 

While one could also consider five-point amplitudes (see \mtabref{tab:five_point}), they do not provide more information, and we will omit that analysis. 
The six-point amplitude considered in this section constrains the contact terms that we previously introduced, such as the unusual one involving the $f_\pm$ factor. Note that once we have fixed the four-point interactions, we cannot introduce any more higher-point interactions in the gauge theory, due to the constraint of renormalizablity. Thus the higher-point calculations will either work out automatically, or the \ghost{} double copy has to be reworked from scratch. 

We first introduce the independent graph topologies. The six-point amplitude for distinguishable massive external scalars, with at least one \ghost{} mediation, can be expressed in terms of the three graphs
\begin{equation} 
\begin{gathered}
\begin{aligned}
&   
\begin{gathered}
    \includegraphics[scale=0.9]{figures/6point_1}
    \end{gathered}
        \\
 &     \hspace{0.4cm}    N_1(\r1,\r2,\gn3,\gn4,\b5,\b6)
    \end{aligned}
\end{gathered}
    \hspace{0.3cm}
    ,
    \hspace{0.3cm}
    \begin{gathered}
        \begin{aligned}
&            \begin{gathered}
    \includegraphics[scale=0.9]{figures/6point_2}
    \end{gathered}
    \\
 &  \hspace{0.4cm} N_2(\r1,\r2,\gn3,\gn4,\b5,\b6)
        \end{aligned}
    \end{gathered}
 \hspace{0.3cm}
 ,
 \hspace{0.3cm}
 \begin{gathered}
     \begin{aligned}
&         \begin{gathered}
    \includegraphics[scale=0.9]{figures/6point_3}
    \end{gathered}
    \\
& \hspace{0.2cm} N_3(\r1,\r2,\gn3,\gn4,\b5,\b6)
     \end{aligned}
 \vspace{0.8cm}
 \end{gathered}
 \hspace{0.2cm}
 ,
\end{equation}
and the corresponding numerator notation that we use throughout this section is displayed below the graphs.

Using kinematic Jacobi-like relations, we can rearrange the gluon interaction and get the following three-term numerator identity
\begin{align}\label{eq:sixpoint jacobi}
N \left(
    \begin{gathered}
   \includegraphics[scale=0.8]{figures/6point_1_s}
    \end{gathered}
    \right)
    &= 
N \left(
     \begin{gathered}
    \includegraphics[scale=0.8]{figures/6point_1}
    \end{gathered}
    \right)
    - 
N \left(
    \begin{gathered}
    \includegraphics[scale=0.8]{figures/6point_1_u}
    \end{gathered}
    \right)
   ,
    \end{align}
  which fully determines $N_3$ in terms of $N_1$. Likewise, using the bonus relation
   \eqref{Eq: Jacobi d2m2}, it follows that $N_3$ is also determined in terms of $N_2$,
\begin{align}\label{Eq: six-point bonus relation}
    \fD \times N\left(
\begin{gathered}
    \includegraphics[scale=0.8]{figures/6point_3}
    \vspace{0.1cm}
    \end{gathered}
    \right)
    &=   
    N \left(
\begin{gathered}
    \includegraphics[scale=0.8]{figures/6point_2}
    \end{gathered}
    \right)
    - 
N \left(
     \begin{gathered}
    \includegraphics[scale=0.8]{figures/6point_2_u}
    \end{gathered}
    \right) .
\end{align}
Using conventional notation, the  functional relations for the numerators are 
\begin{align}
     N_3(\gn3,\gn4,\b5,\b6,\r1,\r2) &= 
     N_1(\r1,\r2,\gn3,\gn4,\b5,\b6)
    -N_1(\r1,\r2,\gn4,\gn3,\b5,\b6),\nn
    \\
      \fD \times N_3(\r1,\r2,\gn3,\gn4,\b5,\b6)  
   &= 
   N_2(\r1,\r2,\gn3,\gn4,\b5,\b6)
     -
    N_2(\r1,\r2,\gn4,\gn3,\b5,\b6)\,,
\end{align}
and to solve these we need to make use of ans\"{a}tze for both $N_1$ and $N_2$.

The ansatz takes the same general form for both numerators (with $i=1,2$)
\begin{multline} \label{N1N2ansatz}
   N_i(\r1,\r2,\gn3,\gn4,\b5,\b6) =
   (\mu_1 \cdot \mu_3) \Big( c_{1}^{(i)} (k_1 \cdot k_2) +...+c_{12}^{(i)} \mu_5^2 \Big)
   + (\mu_1 \cdot \mu_5) \Big( c_{13}^{(i)} (k_1 \cdot k_2) +...+c_{24}^{(i)} \mu_5^2 \Big)\\
   +(\mu_3 \cdot \mu_5) \Big( c_{25}^{(i)} (k_1 \cdot k_2) +...+c_{36}^{(i)} \mu_5^2 \Big)
   + c_{37}^{(i)} (\mu_1 \cdot \mu_3)(\mu_1 \cdot \mu_5)
   \\
 + c_{38}^{(i)} (\mu_1 \cdot \mu_3)(\mu_3 \cdot \mu_5)
   + c_{39}^{(i)} (\mu_1 \cdot \mu_5)(\mu_3 \cdot \mu_5)
  ,
\end{multline}
giving in total $78$ parameters to be fixed. Using automorphism constraints on the three numerators, such as $N_2(\r1,\r2,\gn3,\gn4,\b5,\b6)=N_2(\b6,\b5,\gn4,\gn3,\r2,\r1)$, fixes 42 of the free parameters, the remaining ones will be fixed using cuts.

\subsection{Maximal cuts}

The six-point amplitude must satisfy the appropriate factorization limits, and we will start with the maximal cuts in the gauge theory. To sew together the three-point amplitudes, we make use the \ghost{} state sum \eqref{Eq: ghost state projector}, and for the gluon state sum we insert the state projector $P^{\mu\nu}$ that appeared in \eqn{GRstateDecomp}. The factorizations are then straightforward to work out,
\begin{align}
    \begin{gathered}
   \includegraphics[scale=0.9]{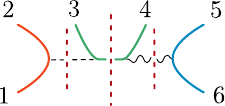}
    \end{gathered}
    =&
    \sum_{\rm states}
    A(\r1, \r2, \g{l}_1)
    \cdots
    A(\b5, \b6, -l_3)
    = (\mu _1\cdot \mu _3 )(k_4\cdot k_6),
    \\
    \begin{gathered}
   \includegraphics[scale=0.9]{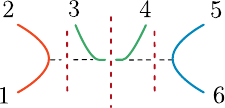}
    \end{gathered}
    =& 
     \sum_{\rm states}
    A(\r1, \r2, \g{l}_1)
    \cdots
    A(\b5, \b6, -\g{l}_3)
    =( \mu _1\cdot \mu _3) ( \mu _3\cdot \mu _5 ),
    \\
    \begin{gathered}
   \includegraphics[scale=0.9]{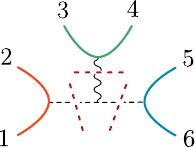}
    \end{gathered}
    =&  \sum_{\rm states}
    A(\r1, \r2, \g{l}_1)
    \cdots
    A(\b5, \b6, -l_3)
    =- (\mu _1\cdot \mu _5) (k_4\cdot k_{56}),
\end{align}
where we use the shorthand $k_{ij}= k_i +k_j$ for sums of momenta.  
We match the factorizations to the numerator ans\"{a}tze  \eqref{N1N2ansatz} and constrain a subset of the free coefficients. Next, we jump directly to the gravitational double copy and verify the construction for the maximal cuts. 

\subsubsection*{Gravity comparison}\label{Sec: 6-point gravity}

The gauge-theory numerators still contain some free parameters associated to contact terms, but we can nevertheless verify the double copy by appropriately squaring the numerators and compare to the gravitational maximal cuts, as obtained from the difference-cut formula~\eqref{eq: dilaton difference}.
For example, consider the partially constrained numerator $N_2(\r1,\r2,\gn3,\gn4,\b5,\b6)$ in the canonical labeling, it is now
\begin{equation}
    N_2 =(\mu _1\cdot \mu _3 )(\mu _3\cdot \mu _5)+ (\mu _1\cdot \mu _5 )\left( (c_5-c_4) k_3\cdot k_{12}+ c_5 k_4\cdot k_{56}+ c_4 k_5\cdot k_{56}\right)\,.
\end{equation}
The corresponding gravitational numerator is obtained from two such copies, with possibly different undetermined parameters, giving  
\begin{align} \label{N2DC}
    \mathcal{N}_2
    =& N_2\tilde{N}_2 \nn\\
    =&
    \Big[(\mu _1\cdot \mu _3 )(\mu _3\cdot \mu _5)+ (\mu _1\cdot \mu _5 )\left( (c_5-c_4) k_3\cdot k_{12}+ c_5 k_4\cdot k_{56}+ c_4 k_5\cdot k_{56}\right) \Big]\nn\\
    &
    \times
    \Big[(\mu _1\cdot \mu _3 )(\mu _3\cdot \mu _5)+ (\mu _1\cdot \mu _5 )\left( (\tilde{c}_5-\tilde{c}_4) k_3\cdot k_{12}+ \tilde{c}_5 k_4\cdot k_{56}+ \tilde{c}_4 k_5\cdot k_{56}\right) \Big],
\end{align}
where $c_i$ and $\tilde{c_i}$ are independent. Similar partially constrained expressions for $\mathcal{N}_1$ and $\mathcal{N}_3$ can be worked out from the ans\"{a}tze \eqref{N1N2ansatz}.

There are two distinct maximal cuts corresponding to the half-ladder and 3-fork topology, and we compute them either from the difference-cut formula~\eqref{eq: dilaton difference} or as a double copy. The free parameters should drop out and the cuts should match. First we consider the maximal cut of the half-ladder as obtained from the difference cut
\begin{align}\label{Eq:6point maxcut1 gravity}
     \left. \begin{gathered}
          \includegraphics[scale=0.9]{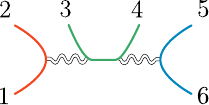}
      \end{gathered} \, \right|_{\Delta\mathfrak{C}}
      & \equiv ~
    \mathfrak{C}^{\perp\!\!\!\perp\!\!\!\perp\!\!\!\perp}_{\text{YM}^2}(\r1,\r2,\gn3,\gn4,\b5,\b6) 
    -\mathfrak{C}^{\perp\!\!\!\perp\!\!\!\perp\!\!\!\perp}_{\text{GR}}(\r1,\r2,\gn3,\gn4,\b5,\b6)\nn\\
&=   \hspace{0.2cm}
    m_1^2 m_3^2 \frac{ \left(k_4\cdot k_6\right){}^2}{D_s - 2}
+
    m_3^2m_5^2 \frac{  \left(k_1\cdot k_3\right){}^2}{D_s - 2}
    -
    \frac{m_1^2 m_3^4 m_5^2}{(D_s - 2)^2}.
\end{align}
To avoid cluttering the pictorial diagram, we indicate that this is a difference cut by the $\Gamma|_{\Delta\mathfrak{C}}$ notation, and all exposed propagators are assumed to be on shell, thus we omit the dashed red cut lines above, as well as in the rest of this paper.   

From the double copy, as given by the numerators ${\cal N}_1$ and ${\cal N}_2$ \eqref{N2DC}, we can compute the same maximal cut which involves the three contributions 
\begin{equation} \label{Eq:6point maxcut1 DC}
    \begin{aligned}
    \left.
      \begin{gathered}
          \includegraphics[scale=0.9]{figures/6point_1_gravity_thick.pdf}
      \end{gathered}
      \, \right|_{\text{DC}}
      \equiv &
    \left(
    \begin{gathered}
    \includegraphics[scale=0.9]{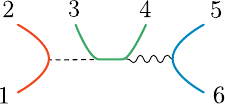}
    \end{gathered}
    \right)^{\hspace{-0.2cm}2}
    +
    \left(
\begin{gathered}
    \includegraphics[scale=0.9]{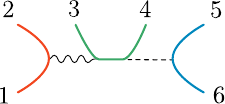}
    \end{gathered}
    \right)^{\hspace{-0.2cm}2}
\\
&
\hspace{4.4cm}-
    \left|
    \begin{gathered}
    \includegraphics[scale=0.9]{figures/6point_2}
    \end{gathered}
    \right|^2
    ,
    \\
    \end{aligned}
\end{equation}
where the restriction $\Gamma|_{\rm DC}$ indicates that the double copy method was used for computing the cut. 
It is a simple exercise to check that the gravitational numerators $\mathcal{N}_1$ and $\mathcal{N}_2$ gives a maximal cut \eqref{Eq:6point maxcut1 DC} that agrees with \eqn{Eq:6point maxcut1 gravity}. Note that in \eqn{Eq:6point maxcut1 DC} it is clear that the two first diagrams give a double counting of contributions with two contiguous dilaton lines, since the squared gluon contains a dilaton, and thus the third diagram comes with the opposite sign to compensate for the overcount.

Considering the maximal cut of the 3-fork topology gives
\begin{align}\label{Eq:6point maxcut2 gravity}
\left.
    \begin{gathered}
        \includegraphics[scale=0.9]{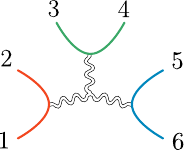}
    \end{gathered}
    \, \right|_{\Delta\mathfrak{C}}
    &\equiv
 \mathfrak{C}^{\,\vdash \hskip-1.4mm {}^{\scriptstyle \top} \hskip-1.5mm \dashv}_{\text{YM}^2}(\r1,\r2,\gn3,\gn4,\b5,\b6) -\mathfrak{C}^{\,\vdash \hskip-1.4mm {}^{\scriptstyle \top} \hskip-1.5mm \dashv}_{\text{GR}}(\r1,\r2,\gn3,\gn4,\b5,\b6)\nn\\
       &= 
    m_1^2 m_5^2 \frac{ \left(k_4\cdot k_{56}\right)^2}{D_s-2}
    +
    m_1^2 m_3^2 \frac{ \left(k_6\cdot k_{34}\right)^2}{D_s-2}
    +
    m_3^2 m_5^2
    \frac{ \left(k_2\cdot k_{56}\right)^2}{D_s -2}.
    \end{align}
and we compare this difference cut with the cut of the double copy that involves three permutations of $\mathcal{N}_3$,
\begin{equation}
   \begin{aligned}
     \left.\begin{gathered}
        \includegraphics[scale=0.87]{figures/6point_3_gravity_thick.pdf}
    \end{gathered}\,\right|_{\text{DC}}
            \!\!\! &\equiv \!
        \left(
    \begin{gathered}
    \includegraphics[scale=0.87]{figures/6point_3}
    \end{gathered}
     \right)^{\hspace{-0.2cm}2}
    \!+\!
    \left(
     \begin{gathered}
    \includegraphics[scale=0.87]{figures/6point_1_s}
    \end{gathered}
     \right)^{\hspace{-0.2cm}2}
\!+\!
\left(
   \begin{gathered}
    \includegraphics[scale=0.87]{figures/6point_3_flip}
    \end{gathered}
     \right)^{\hspace{-0.2cm}2}\,,
    \end{aligned}
\end{equation}
and we find perfect agreement. Next, we need to verify and constrain the contact terms.

\subsection{Next-to-maximal gravity cuts}

We now use the GR information at next-to-maximal-cut (NMC) level to fully fix the free parameters multiplying the contact terms of the gauge theory numerators. There are two distinct NMCs that one can compare to GR; however, one of these trivially  works out and thus contains no new information. Hence, we will focus only on the non-trivial one that constrains our construction. This difference cut is given by the expression 
\begin{align}  \label{NMCfromDifCut}
\left.
\begin{gathered}
    \includegraphics[scale=0.8]{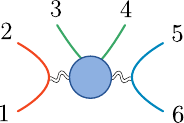}
    \end{gathered} 
    \right|_{\Delta\mathfrak{C}}
\equiv&~ 
\mathfrak{C}^{\rm NMC}_{\rm YM^2}(\r1,\r2,\gn3,\gn4,\b5,\b6)  - \mathfrak{C}^{\rm NMC}_{\rm GR}(\r1,\r2,\gn3,\gn4,\b5,\b6)
\nn\\
    =& \hspace{0.4cm}
\frac{1}{D_s-2} \Bigg[
\frac{1}{(k_{456}^2 - m_3^2)}
\left(\frac{1}{4}
m_1^2 m_3^2t_{45}^2
+
\frac{1}{4}
m_5^2m_3^2 t_{23}^2
- 
\frac{m_1^2 m_3^4 m_5^2}{D_s-2}
\right)
\nn\\
&+
\frac{1}{(k_{356}^2-m_3^2)}
\left(\frac{1}{4}
m_1^2 m_3^2 t_{35}^2
+
\frac{1}{4}
m_3^2 m_5^2 t_{24}^2
-\frac{m_1^2 m_3^4 m_5^2}{ D_s-2 }
\right)
\nn\\
&+
\frac{1}{s_{34}}
\Big(
m_1^2m_5^2 \left(k_3\cdot k_{56}\right)^2
+m_1^2 m_3^2\left(k_5\cdot k_{34}\right)^2
+ m_5^2 m_3^2\left(k_2\cdot k_{56}\right)^2
\Big)
\nn\\
&+\frac{m_1^2 m_5^2 \left(k_{356}^2-m_3^2\right)}{4}-\frac{m_1^2 m_3^2 m_5^2}{D_s-2}
\Bigg]
,
\end{align}
where the exposed propagators are on shell, sums of momenta are $k_{i\ldots j}= k_i+\ldots+k_j$ and we recall that $t_{ij}=2 k_i\cdot k_j$. 
%
The same cut in terms of the double-copy graphs takes the form
\begin{align} \label{NMCfromDC}
\left.
    \begin{gathered}
    \includegraphics[scale=0.8]{figures/6point_gravity_n1cut_thick.pdf}
    \end{gathered}
    \right|_{\text{DC}} \hspace{-0.3cm}
    \equiv& 
    \left(
    \begin{gathered}
   \includegraphics[scale=0.8]{figures/6point_1.pdf}
    \end{gathered}
    \right)^{\hspace{-0.2cm}2}
    +
    \left(
    \begin{gathered}
   \includegraphics[scale=0.8]{figures/6point_1_flip.pdf}
    \end{gathered}        \right)^{\hspace{-0.2cm}2}
    +
    \left(
    \begin{gathered}
   \includegraphics[scale=0.8]{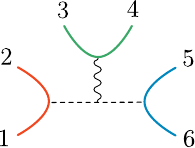}
    \end{gathered}
    \right)^{\hspace{-0.2cm}2}
    \nn\\
&\hspace{-0.4cm} +
\left(
    \begin{gathered}
   \includegraphics[scale=0.8]{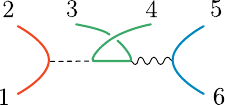}
    \end{gathered}
    \right)^{\hspace{-0.2cm}2}
    +
    \left(
    \begin{gathered}
   \includegraphics[scale=0.8]{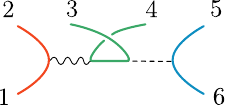}
    \end{gathered}
    \right)^{\hspace{-0.2cm}2}
    +
    \left(
    \begin{gathered}
   \includegraphics[scale=0.8]{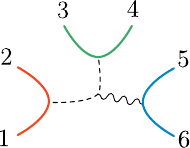}
    \end{gathered}
    \right)^{\hspace{-0.2cm}2}
\nn \\
&\hspace{-0.4cm} 
    -\,
    \left|
    \begin{gathered}
   \includegraphics[scale=0.8]{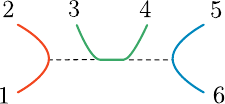}
    \end{gathered}
 \right|^2
     -~
     \left|
    \begin{gathered}
   \includegraphics[scale=0.8]{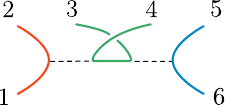}
    \end{gathered}
    \right|^2
    +~
\left(
     \begin{gathered}
   \includegraphics[scale=0.8]{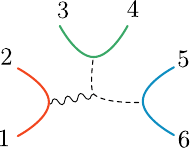}
    \end{gathered}
    \right)^{\hspace{-0.2cm}2}\!,
    \end{align}
where off-shell propagator connecting to legs 3 and 4 is included in the diagrams on the right-hand side. 

The unfixed parameters \eqref{N2DC} of the numerators now contribute, and matching  \eqn{NMCfromDifCut} with \eqn{NMCfromDC} gives a non-trivial constraint.  We solve for this constraint, and find that the contact terms consolidate such that they appear in the  following conjugate numerators of the second graph topology:
\begin{align}
N_2(\r1,\r2,\gn3,\gn4,\b5,\b6) 
=
~(\mu _1\cdot \mu _3)( \mu _3\cdot \mu _5)+\frac{-1 + \sqrt{D_s-1 }}{4} (\mu _1\cdot \mu _5) \left(k_{456}^2
    - m_3^2\right), \nn \\
\tilde{N}_2(\r1,\r2,\gn3,\gn4,\b5,\b6) 
=
~(\mu _1\cdot \mu _3)( \mu _3\cdot \mu _5)+\frac{-1 - \sqrt{D_s-1 }}{4} (\mu _1\cdot \mu _5) \left(k_{456}^2
    - m_3^2\right)\,,
\end{align}
which only differs by the sign choice in front of the square root.
The remaining numerators do not depend on this sign, and they are
\begin{align} \label{N1N3solution}
N_1(\r1,\r2,\gn3,\gn4,\b5,\b6) 
=&
~\frac{1}{2}  (\mu _1\cdot \mu _3)\left(k_4\cdot k_6-k_4\cdot k_5\right), \nn \\
N_3(\r1,\r2,\gn3,\gn4,\b5,\b6)
=&
~\frac{1}{2} (\mu _1\cdot \mu _5)  \left(k_3\cdot k_{56}-k_4\cdot k_{56}\right) ,
\end{align}
and the conjugate numerators are the same $\tilde{N}_i= N_i$.
Note that the Jacobi-like relation in \eqn{eq:sixpoint jacobi} gives $N_3$ from $N_1$. 

The bonus relation in \eqn{Eq: six-point bonus relation} can now be checked and we can fix the $\fD$ factor, which prior to the six-point calculation is unknown. The bonus relation evaluates to
\begin{equation}
\begin{aligned}
    N_3(\r1,\r2,\gn3,\gn4,\b5,\b6) =& 
    ~
    \frac{1}{\fDp} ~\Big[ N_2(\r1,\r2,\gn3,\gn4,\b5,\b6) - N_2(\r1,\r2,\gn4,\gn3,\b5,\b6) \Big]
    \\
    =& ~
    \frac{1}{\fDp}~\frac{-1 + \sqrt{D_s-1}}{2}
     ( \mu _1\cdot \mu _5)\left(k_3\cdot k_{56}-k_4\cdot k_{56}\right)
   .
\end{aligned}
\end{equation}
Comparing to $N_3$ in \eqn{N1N3solution} gives that
\begin{equation} \label{fsolution}
    \fD = -1 \pm\sqrt{D_s-1}\,,
\end{equation}
where the $\pm$ follows from repeating the same exercise for the $\tilde N_i$ numerators. 

Since all the contact terms are now fixed, we can check that we obtain the correct N${}^2$MC difference cuts from the double copy, and likewise after comparing the complete six-point amplitude $M(\r1,\r2,\gn3,\gn4,\b5,\b6)$ to the results in \rcite{Carrasco:2020ywq} we find a perfect match. 

\subsection*{Final expressions}
 The final expressions for the six-opoint numerators, in canonical labeling $(\r1,\r2,\gn3,\gn4,\b5,\b6)$, are summarized here 
 \begin{align}
   N_1 &=  
N\left(
   \begin{gathered}
   \includegraphics[scale=0.8]{figures/6point_1}
    \end{gathered}
    \right)
    =\frac{1}{2}  (\mu _1\cdot \mu _3)\left(k_4\cdot k_6-k_4\cdot k_5\right),
    \\
N_2 &= 
N\left(
\begin{gathered}
  \includegraphics[scale=0.8]{figures/6point_2}
    \end{gathered}
    \right)= (\mu _1\cdot \mu _3)( \mu _3\cdot \mu _5)+\frac{\fD}{4} (\mu _1\cdot \mu _5) \left(k_{456}^2
    - m_3^2\right)
    ,\\
N_3
   &=N
   \left(
   \begin{gathered}
     \includegraphics[scale=0.8]{figures/6point_3}
    \end{gathered}
   \right)
   =~\frac{1}{2} (\mu _1\cdot \mu _5)  \left(k_3\cdot k_{56}-k_4\cdot k_{56}\right).
    \end{align}
     The corresponding gravitational numerators are 
\begin{align}
    \mathcal{N}_1 &= N_1^2 =  \frac{1}{2}m_1^2m_3^2 \left(k_4\cdot k_6-k_4\cdot k_5\right)^2,\\
    \mathcal{N}_2 &=N_2 \tilde{N}_2=
   \frac{D_s-2}{16}m_1^2m_5^3  (k_{456}^2-m_3^2)^2 - m_1^2m_3^2m_5^2\left( m_3^2 + \frac{1}{2}(k_{456}^2-m_3^2)\right),\\
    \mathcal{N}_3 &= N_3^2=\frac{1}{4} m_1^2m_5^2 \left(k_3\cdot k_{56}-k_4\cdot k_{56}\right)^2.
\end{align}


\subsubsection*{Yang-Mills amplitude}

For completeness, we also give the six-point numerators that are purely mediated by gluons.  They are given by the following two Yang-Mills-scalar diagrams that satisfy color-kinematics duality: 
\begin{align}
    N
    \left(
    \begin{gathered}
        \includegraphics[scale=0.8]{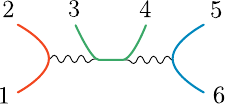}
    \end{gathered}
    \right) =
\frac{1}{8} t_{456} \left( \frac{1}{2} t_{24}+t_{25}- 3 k_2\cdot k_{456}+\frac{1}{2}t_{46}-\frac{1}{2}t_{36}\right)
\nn\\
+\frac{1}{16} \left(4 k_2\cdot k_{456}+s_{12}\right) \left(2t_{46}+s_{56}\right)\nn\\
+ \frac{1}{32} t_{456}\left(t_{356}-t_{456}-2s_{12}\right),
\\
     N
    \left(
    \begin{gathered}
        \includegraphics[scale=0.8]{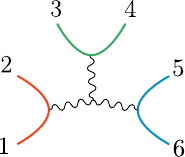}
    \end{gathered}
    \right) = \frac{1}{8} s_{12} \left(\frac{1}{2}t_{46}-\frac{1}{2}t_{45}+k_6\cdot k_{34}+k_2\cdot k_{56}-2 k_2\cdot k_{45}\right)
    \nn\\
    +\frac{1}{8}s_{56} \left(\frac{1}{2}t_{46}-\frac{1}{2}t_{45}-k_6\cdot k_{34}-k_2\cdot k_{56}+2 k_2\cdot k_{45}\right)
    \nn\\
    +\frac{1}{8} s_{34} \left(\frac{1}{2}t_{45}-\frac{1}{2}t_{46}+k_6\cdot k_{34}-k_2\cdot k_{56}+2 k_2\cdot k_{45}\right)
    \nn\\
    + \frac{1}{2}t_{24}(k_6\cdot k_{34})-\frac{1}{2}t_{45}(k_2\cdot k_{56})+\frac{1}{2}t_{25}( k_4\cdot k_{56})
    \nn\\
    -\frac{1}{32}\left(s_{12}^2-s_{34}^2+s_{56}^2\right)+\frac{s_{12}s_{56}}{16}, 
 \end{align}
where we use shorthand notation for the inverse propagators $t_{356} = (k_3+k_5+k_6)^2-m_3^2$ and $t_{456} = (k_4+k_5+k_6)^2-m_3^2$. 
Using the double-copy prescription in \eqn{Eq: Gravity amp prescription} the corresponding GR amplitude $M_{\rm GR}(\r1,\r2,\gn3,\gn4,\b5,\b6)$ can be worked out. The necessary ingredients in the formula are the YM and \ghost{} numerators given in this section.

\section{One-loop amplitude}\label{Sec:one-loop}

In this section, we bootstrap the integrand of the four-point one-loop amplitude with external massive scalars and internal \ghost{} states. The strategy is the same as at tree level --- we write down all relevant cubic graph topologies, use numerator relations to determine a minimal set of unknown numerators and give ans\"{a}tze to these. The free parameters of the ans\"{a}tze are fixed using automorphism properties, unitarity cuts, kinematic Jacobi and bonus relations. We then verify the result by comparing the double copy to the difference cuts \eqref{eq: dilaton difference} using input from GR \cite{Carrasco:2021bmu}. At the one-loop level, we encounter for the first time unwanted contributions from the axion (or antisymmetric tensor $B^{\mu \nu}$ in general dimension) which appear in the bubble graphs. We will remove these contributions by hand. 

\begin{table}[]
    \centering
     \begin{tabular}{
     M{0.09\linewidth}|
    M{0.21\linewidth}
    |M{0.7\linewidth}}
    $N_i $ & \text{Topology} & \text{Numerator}
 \\
 \hline
 $N_1 $ &
 \vspace{-0.5cm}
\begin{gather*}
     \begin{gathered}
    \includegraphics{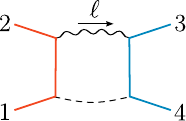} 
    \end{gathered}
\end{gather*}
 \vspace{-0.5cm}
    &    
     \vspace{-0.5cm}
    \begin{gather*}
         - (\mu _1\cdot \mu _3) \left(k_1\cdot k_4+\frac{1}{4}\ell^2\right)
    \end{gather*}
     \vspace{-0.5cm}
    \\
$N_2 $ &
 \vspace{-0.5cm}
    \begin{gather*}
        \begin{gathered}
    \includegraphics{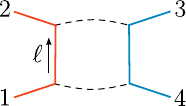}
    \end{gathered}
    \end{gather*} 
\vspace{-0.5cm}
    &  
     \vspace{-0.3cm}
    \begin{gather*}
          \left( \mu_1^2 -\frac{\fD}{4}(\ell^2-\mu_1^2) \right)
 \left( \mu_3^2 -\frac{\fD}{4}((\ell+k_{14})^2-\mu_3^2) \right)
     \end{gather*}
      \vspace{-0.3cm}
\\
$N_3 $ &
 \vspace{-0.5cm}
\begin{gather*}
    \begin{gathered}
    \includegraphics{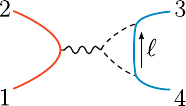}
\end{gathered}
\end{gather*}
 \vspace{-0.5cm}
    & 
    \vspace{-0.3cm}
\begin{gather*}
\frac{1}{2}~(k_2-k_1)\cdot (\ell+ k_4)  \left(   \mu_3^2 
    - \frac{\fD}{4}\left(\ell^2-\mu_3^2\right)\right)
\end{gather*}
\vspace{-0.3cm}
    \\
$N_4 $ &
\vspace{-0.5cm}
    \begin{gather*}
        \begin{gathered}
    \includegraphics{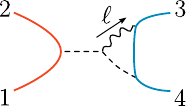}
    \end{gathered}
    \end{gather*}    
    \vspace{-0.5cm}
    & 
    \vspace{-0.5cm}
   \begin{gather*}
-    \frac{1}{2} ~(\mu _1\cdot \mu _3) \left((k_3+k_4)^2-\ell^2\right)
   \end{gather*}
   \vspace{-0.5cm}
    \\
    $N_5 $ &
    \vspace{-0.5cm}
     \begin{gather*}
         \begin{gathered}
         \includegraphics{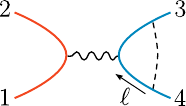}
     \end{gathered}
     \end{gather*}
         \vspace{-0.5cm}
    & 
    \vspace{-0.5cm}
    \begin{gather*}
        \frac{1}{2} ~\mu_3^2 ~(k_2- k_1)\cdot \ell
    \end{gather*}
    \vspace{-0.5cm}
    \\
    $N_6 $ &
    \vspace{-0.5cm}
    \begin{gather*}
        \begin{gathered}
        \includegraphics{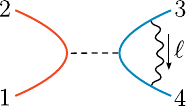}
    \end{gathered}
    \end{gather*}
        \vspace{-0.5cm}
    &  
    \vspace{-0.5cm}
    \begin{gather*}
        ( \mu _1\cdot \mu _3) \left( k_3\cdot k_4+\frac{1}{4}\ell^2\right)
    \end{gather*}
    \vspace{-0.5cm}
    \\
    $N_7 $ &
    \vspace{-0.5cm}
    \begin{gather*}
        \begin{gathered}
    \includegraphics{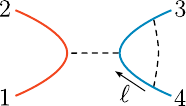}
    \end{gathered}
    \end{gather*}
    \vspace{-0.5cm}
    & 
    \vspace{-0.5cm}
    \begin{gather*}
        -\mu_3^2 \left(\mu _1\cdot \mu _3\right)
    \end{gather*}
    \vspace{-0.5cm}
    \end{tabular}
\caption{The box and triangle one-loop graph topologies with kinematic numerators $N_i$ in the canonical labeling $N_i(\r1, \r2, \b3, \b4; \ell)$. The numerators were constrained on the physical unitarity cuts, and any ans\"{a}tze parameters not fixed by either cuts or numerator relations have been set to zero.}
    \label{tab:one-loop topos}
\end{table}

  \begin{table}[]
    \centering
     \begin{tabular}{
     M{0.09\linewidth}|
    M{0.21\linewidth}
    |M{0.7\linewidth}}
    $N_i $ & \text{Topology} & \text{Numerator}
 \\
 \hline
 $N_8 $ &
 \vspace{-0.5cm}
 \begin{gather*}
     \begin{gathered}
    \includegraphics{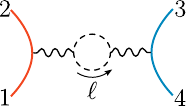}     
 \end{gathered}
 \end{gather*}
 \vspace{-0.5cm}
        & 
        \begin{gather*}
            -\frac{1}{4}~(k_3-k_4)\cdot \ell~(k_2-k_1)\cdot \ell
        \end{gather*}
\\
 $N_9 $ &
\vspace{-0.5cm}
        \begin{gather*}
            \begin{gathered}
            \includegraphics{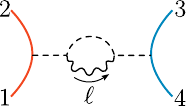}
        \end{gathered}
        \end{gather*}
        \vspace{-0.5cm}
     &
\begin{gather*}
    (\mu _1\cdot \mu _3)\left((k_4+k_3)^2-\ell^2\right)
\end{gather*}
  \\
   $N_{10} $ &
  \vspace{-0.5cm}
    \begin{gather*}
        \begin{gathered}
        \includegraphics{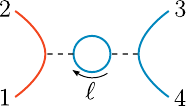}
    \end{gathered}
    \end{gather*}
    \vspace{-0.5cm}
    &
\begin{gather*}
     -\mu_3^2 \left(\mu _1\cdot \mu _3\right)
\end{gather*}
    \\
 $N_{11} $ &    
\vspace{-0.5cm}
    \begin{gather*}
        \begin{gathered}
        \includegraphics{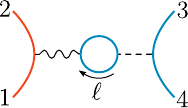}
    \end{gathered}
    \end{gather*}
    \vspace{-0.5cm}
    &  
    \begin{gather*}
        \frac{1}{2} ~\mu_3^2~ ( k_1-k_2)\cdot \ell
    \end{gather*}
\\
 $N_{12} $ &
\vspace{-0.5cm}
    \begin{gather*}
        \begin{gathered}
        \includegraphics{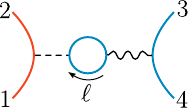}
    \end{gathered}
    \end{gather*}
    \vspace{-0.5cm}
    &
\begin{gather*}
   \frac{1}{2}( \mu _1\cdot \mu _3 )(k_3-k_4)\cdot \ell
\end{gather*}
    \end{tabular}
    \caption{The bubble one-loop graph topologies with kinematic numerators $N_i$ in the canonical labeling $N_i(\r1, \r2, \b3, \b4; \ell)$. The bubble numerators are fully determined by eqs.~\eqref{N8relation}--\eqref{N12relation}.}
    \label{tab:one-loop topos 2}
\end{table}

The one-loop four-point amplitude can be expressed in terms of the twelve graph topologies listed in \mtabref{tab:one-loop topos}--\ref{tab:one-loop topos 2}, where each graph is paired with its kinematic numerator. The numerators are related using the same kinematic Jacobi and bonus relations as discussed at tree level, from which one can work out the following minimal set of five unknown one-loop numerators: 
\begin{equation}\label{Eq: one-loop basis graphs}
\begin{aligned}
\begin{gathered}
    \begin{gathered}
    \includegraphics[scale=0.9]{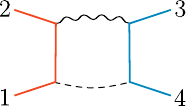} 
    \end{gathered}
    ~
    ,
    ~
      \begin{gathered}
    \includegraphics[scale=0.9]{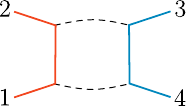} 
    \end{gathered}
    ~
    ,
    ~
        \begin{gathered}
    \includegraphics[scale=0.9]{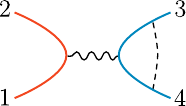}
    \end{gathered}
    ~
    ,
    ~
\\
N_1(\r1,\r2,\b3,\b4;\ell) \hspace{0.8cm} 
N_2 (\r1,\r2,\b3,\b4;\ell) \hspace{0.8cm} 
N_5(\r1,\r2,\b3,\b4;\ell)
    \\
    \begin{gathered}
    \includegraphics[scale=0.9]{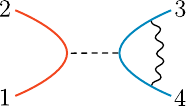}
    \end{gathered}
        ~
    ,
    ~
     \begin{gathered}
    \includegraphics[scale=0.9]{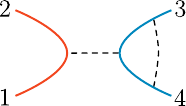}
    \end{gathered}
       ~
    .
    \\
 N_6(\r1,\r2,\b3,\b4;\ell) \hspace{0.8cm}  
 N_7(\r1,\r2,\b3,\b4;\ell)
    \end{gathered}
    \end{aligned}
\end{equation}
 
The numerators of the remaining topologies can then be expressed as linear combinations of these.  As an example, consider the $N_3$ numerator in \mtabref{tab:one-loop topos}, which is determined by the following bonus relation:
\begin{equation}\label{Eq: triangle Jacobi}
\fD \times N\left(
 \begin{gathered}
\includegraphics[scale=0.9]{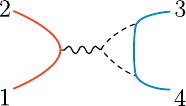}
    \end{gathered}
    \right)
 =
 N \left(
     \begin{gathered}
\includegraphics[scale=0.9]{figures/box2.pdf}
    \end{gathered}
    \right)
    -
N \left(
\begin{gathered}
\includegraphics[scale=0.9]{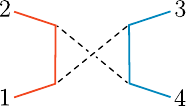}
\end{gathered}   
\right)
,
\end{equation}
which follows from applying \eqn{Eq: Jacobi d2m2} to the subgraphs centered on the leftmost internal propagators. Expressing the relation \eqref{Eq: triangle Jacobi} on conventional functional form gives 
\begin{equation}
    N_3(\r1,\r2,\b3,\b4;\ell) = \frac{1}{\fD} \Big( N_2(\r1,\r2,\b3,\b4;k_2+k_3-\ell)-N_2(\r2,\r1,\b3,\b4;k_1+k_3-\ell) \Big)\,,
\end{equation}
where the loop momentum in the last argument is non-trivially shifted on the right-hand side in order to match the canonical labeling of each graph. 

The remaining numerators in \mtabref{tab:one-loop topos}--\ref{tab:one-loop topos 2} can be related to the ones in \eqn{Eq: one-loop basis graphs} through the following kinematic Jacobi or bonus relations:
\begin{align}
N_4 =&~ N_1(\r1,\r2,\b3,\b4;\ell)-N_1(\r2,\r1,\b3,\b4;\ell),\\
\label{N8relation}
N_8 =& ~\frac{1}{\fD^2} \Big( N_2(\r1,\r2,\b3,\b4;k_2-\ell)-N_2(\r1,\r2,\b4,\b3;k_2-\ell)\nn\\
&\hspace{0.7cm}
-N_2(\r2,\r1,\b3,\b4;k_1-\ell)
+N_2(\r2,\r1,\b4,\b3;k_1-\ell) \Big),\\
N_9 =&~  N_1(\r1,\r2,\b3,\b4;\ell)-N_1(\r1,\r2,\b4,\b3;\ell)
 -N_1(\r2,\r1,\b3,\b4;\ell)+N_1(\r2,\r1,\b3,\b4;\ell),\\
N_{10}=& ~N_{7}(\r1,\r2,\b3,\b4;\ell),\\
N_{11}=& ~N_{5}(\r1,\r2,\b3,\b4;\ell),\\
\label{N12relation}
N_{12}=& ~N_{6}(\r1,\r2,\b3,\b4;\ell)-N_{6}(\r1,\r2,\b4,\b3;\ell)
\end{align}
where the suppressed arguments on the left-hand sides are in the canonical ordering $(\r1,\r2,\b3,\b4;\ell)$.

The corresponding gravitational numerators are double copies of the gauge-theory numerators,
\begin{equation}
    \mathcal{N}_i(\r1,\r2,\b3,\b4;\ell) = N_i(\r1,\r2,\b3,\b4;\ell) \times \tilde{N}_i(\r1,\r2,\b3,\b4;\ell),
\end{equation}
where $N_i$ and $\tilde{N}_i$ have opposite signs for the $\fD$ factor. 

\subsubsection*{Ans\"{a}tze}

The numerators of the twelve graph topologies are now expressed in terms of five unknown numerators in \eqn{Eq: one-loop basis graphs}, and we now introduce ans\"{a}tze for the latter.  
The numerators have to be constructed from the following set of independent dot products:
\begin{equation}
\beta^L_i = 
    \left\{k_1{\cdot} k_2,
    ~k_1{\cdot} k_4,
    ~k_2{\cdot}\ell,
    ~k_3{\cdot} \ell,
    ~k_4{\cdot} \ell,
    ~\ell ^2,
    ~\mu _1^2,
    ~\mu _1{\cdot} \mu _3,
    ~\mu _3^2\,\right\}.
\end{equation}
The most general local ansatz is quadratic in the dot products, giving 45 terms for each of the five unknown numerators,
\begin{equation}
\begin{aligned}
   N_i =& \sum_{1\le j\le k\le 9} c_{jk}^{(i)} \beta_j^L \beta_k^L.
     \end{aligned}
 \end{equation}
 
To fix the $c_{jk}^{(i)}$ parameters we systematically compare to the unitarity cuts of the amplitude, starting with the maximal cuts. We then successively release cut conditions to obtain the NMCs and N${}^2$MCs. We will perform the following quadruple, triple and two-particle cut in detail: 
 \begin{equation}\label{Eq:max cut method}
    \begin{gathered}
    \includegraphics{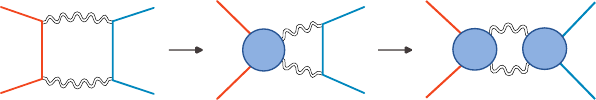}\,, 
    \end{gathered}
\end{equation}
where blobs represent four-point amplitudes and exposed propagators are on-shell legs. For completeness, but omitting the details, we will also consider maximal cuts of triangle and bubble graphs, as well as physical N${}^k$MCs not displayed above. We will not perform unphysical cuts that probe terms of the integrand that integrates to zero, such as tadpoles and bubbles on external legs.

\subsection{Quadruple cuts }

There are two box topologies in the \ghost-mediated one-loop four point amplitude and we can analyse the maximal cuts for these separately. Similar to the previous section, we will omit the red dashed ``cutting"-lines in graphs.  Any exposed line is assumed to be on shell, and furthermore we label the on-shell internal lines by their momenta $\ell_i$, and momentum conservation is left implicit.

\subsubsection*{Gauge theory maximal cuts}

Using the three-point amplitudes defined in \eqns{Eq: 3-point Higgs gluon}{Eq: 3-point mass gluon}, the maximal cut of the box with a single \ghost{} propagator is easily found,
\begin{align}\label{Eq:boxcut1}
    \begin{gathered}
    \includegraphics{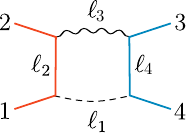} 
    \end{gathered}
    &= \sum_{\text{states}} 
    A(-\g{\ell}_1,\r1, \r{\ell}_2)
    A(\ell_3,-\r{\ell}_2, \r2)
    A(-\ell_3, \b3, \b{\ell}_4)
    A(\g{\ell}_1,-\b{\ell}_4, \b4),
    \nn\\
    &= -(\mu_1 \cdot \mu_3)(k_1 \cdot k_4)\,.
\end{align}
As before, we compare this cut result to the numerator $N_1(\r1, \r2, \b3, \b4; \ell)$, fixing a subset of the ansatz parameters. Similarly, the box with two \ghost-propagators has a maximal cut given by
\begin{align}\label{Eq:boxcut2}
     \begin{gathered}
    \includegraphics[scale=1.0]{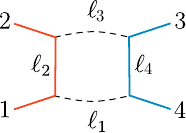} 
    \end{gathered}
    &= 
    \sum_{\text{states}} 
    A(-\g{\ell}_1,\r1, \r{\ell}_2)
    A(\g{\ell}_3,-\r{\ell}_2,\r2) 
    A(-\g{\ell}_3,\b3, \b{\ell}_4)
    A(\g{\ell}_1,-\b{\ell}_4,\b4)
\nn \\
    &= (\mu_1 \cdot \mu_3)^2\,.
\end{align}
 Again, comparing this cut to the numerator  $N_2(\r1, \r2, \b3, \b4; \ell)$ fixes a subset of the ansatz parameters. 
 Considering the maximal cuts of the remaining triangle and bubble topologies, fixes further parameters.

\subsubsection*{Gravity comparison}

The gravitational difference cuts at one loop are defined in the same way as tree level \eqref{eq: dilaton difference} -- they are obtained by subtracting the GR amplitudes obtained in \rcite{Carrasco:2021bmu} from the naive double copy of YM coupled to massive scalars. The difference maximal cut for the box diagram is 
\begin{equation}\label{Eq: Box gravity max cut}
    \begin{aligned}
    \left.
\begin{gathered}
    \includegraphics[scale=1.0]{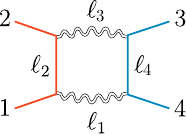}
\end{gathered}
\right|_{\Delta \mathfrak{C}}
\equiv &~\mathfrak{C}_{\text{YM}^2}(\r1,\r2,\b3,\b4;\ell)  -  \mathfrak{C}_{\text{GR}}(\r1,\r2,\b3,\b4;\ell)\\
     =& ~\frac{2}{D_s-2}m_1^2 m_3^2 \left(k_1 \cdot k_4\right)^2 - \frac{1}{(D_s-2)^2} (m_1^2 m_3^2)^2.
    \end{aligned}
\end{equation}
We compare this result to the double copy of the \ghost-amplitude numerators, giving the sum or difference of three contributions
\begin{equation}
\left.
\begin{gathered}
    \includegraphics[scale=0.9]{figures/boxcut_gravity_thick.pdf}
\end{gathered}
\right|_{\text{DC}}
\equiv
    \left(
       \begin{gathered}
    \includegraphics[scale=0.9]{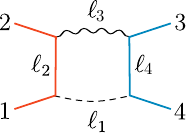} 
    \end{gathered}
    \right)^{\hspace{-0.2cm}2}
        +
    \left(
     \begin{gathered}
     \includegraphics[scale=0.9]{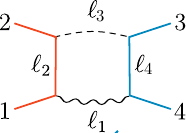}
     \end{gathered}
     \right)^{\hspace{-0.2cm}2}
     -
        \left|
       \begin{gathered}
    \includegraphics[scale=0.9]{figures/boxcut2.pdf} 
    \end{gathered}
    \right|^2,
\end{equation}
where the squaring operation is defined in \eqn{Eq: 2* definition}. We find that the double copy agrees with \eqn{Eq: Box gravity max cut}. In the same vein, we perform maximal cuts of the double copy of the triangle and bubble graphs and find full agreement with the corresponding difference cuts.

\subsection{Triple cuts (NMCs)}

A majority of the NMCs of the one-loop amplitude are automatically satisfied by our numerators once maximal cuts are satisfied. There are, however, contact terms that appear in certain cuts. These contact terms, as we have already seen at tree level, are intimately related with the kinematic Jacobi and bonus relations that constrain our system of numerators. The special bonus relation introduced in \eqn{Eq: Jacobi d2m2} has so far been tested at tree level, and we wish to investigate its behavior at one loop. 

We express the gauge theory triangle cut in terms of the following planar graphs:
\begin{equation}\label{eq:triangle cut}
\begin{aligned}
\begin{gathered}
\includegraphics[scale=1]{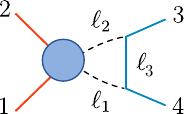}
\end{gathered}  
&= 
        \begin{gathered}
\includegraphics[scale=1.0]{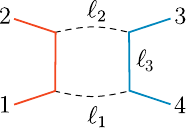}
    \end{gathered}
    +
     \begin{gathered}
\includegraphics[scale=1.0]{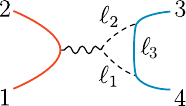}
    \end{gathered}
    \end{aligned}
\end{equation}
where, for clarity, all the on-shell propagators are labeled ($\ell_1$, $\ell_2$ and $\ell_3$), and unlabeled propagators are off shell.  This should be compared to the same cut as obtained by sewing the product of tree amplitudes,
\begin{equation}
\begin{aligned}
  \begin{gathered}
\includegraphics[scale=1]{figures/trianglecut.pdf}
\end{gathered}  
 &=
 \sum_{\text{states}}
    A(\r1,\r2, \g{\ell}_2 ,\g{\ell}_1) ~A(-\g{\ell}_2 ,\b3, \b{\ell}_3 )
    ~A(-\b{\ell}_3, \b4 ,-\g{\ell}_1)\\
&=
    \frac{m_1^2m_3^2}{t_{1\ell}}-\frac{1}{2} \frac{m_3^2t_{1\ell}}{s_{12}}\pm m_3^2\frac{\sqrt{D_s -1}}{4},
\end{aligned}
\end{equation}
where we use the shorthand notation $t_{1\ell} =2 k_1 \cdot \ell_1 = (k_1+\ell_1)^2-m_1^2$. The square root in the cut clearly originates from the contact term that involves $\fD$. We compare this expression with \eqn{eq:triangle cut} and fix some of the remaining ansatz parameters in terms of the $\fD$ factor. To verify that the resulting numerators have the correct contact terms, we again compare with the gravitational difference cut. 

\subsubsection*{Gravity comparison}

The gravitational difference cut now probe non-trivial contact terms. As above, we consider the triple cut given by
\begin{align}\label{eq:NMC cut 1-loop gravity}
\left.
\begin{gathered}
    \includegraphics[scale=1.0]{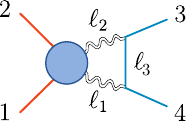}
    \end{gathered}
    \right|_{\Delta \mathfrak{C}}
    \equiv &~
       \mathfrak{C}_{\text{YM}^2}^{\text{NMC}}(\r1,\r2,\b3,\b4;\ell)  -  \mathfrak{C}_{\text{GR}}^{\text{NMC}}(\r1,\r2,\b3,\b4;\ell)
   \nn \\
    =&
    -\frac{1}{(D_s-2)^2}
    \left(\frac{m_1^4 m_3^4}{t_{2\ell}}+\frac{m_1^4 m_3^4}{t_{1\ell}}
    \right)  \nn\\ &
  + \frac{1}{2} \frac{1}{D_s-2}
    \Big(\frac{m_1^2 m_3^2 \left(s_{12}+t_{24}\right){}^2}{t_{1\ell}}+\frac{m_1^2 m_3^2 t_{24}^2}{t_{2\ell}}+\frac{1}{2}\frac{m_3^4 t_{1\ell}^2}{s_{12}}\Big)
   \nn \\ &- \frac{1}{(D_s-2)^2}m_1^2 m_3^4 + \frac{1}{2}\frac{1}{D_s-2} \left( 
    \frac{1}{2} m_3^4 t_{1\ell}+ m_1^2 m_3^2 s_{12}\right)
,
\end{align}
where  $t_{2\ell} =2 k_2 \cdot \ell_1 = (k_2 + \ell_1)^2 - m_1^2$. The last line indeed contains several non-trivial contact terms -- one proportional to $(D_s-2)^{-2}$ and another proportional to $(D_s-2)^{-1}$.

We compare difference cut \eqref{eq:NMC cut 1-loop gravity} to the double copy of the \ghost-graphs,
\begin{equation}
    \begin{aligned}
    \left.
             \begin{gathered}
    \includegraphics[scale=0.8]{figures/trianglecut_gravity_thick.pdf}
    \end{gathered}
    \right|_{\text{DC}}
    \hspace{-0.2cm}
    \equiv&~
   -
        \left|
       \begin{gathered}
    \includegraphics[scale=0.8]{figures/trianglecut_1.pdf} 
    \end{gathered}
    \right|^2
    \hspace{0.1cm}
   -
   \hspace{0.2cm}
    \left|
    \begin{gathered}
        \includegraphics[scale=0.8]{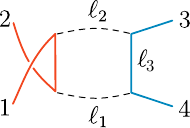}
    \end{gathered}
    \right|^2
-
\hspace{0.2cm}
    \left|
    \begin{gathered}
        \includegraphics[scale=0.8]{figures/trianglecut_2.pdf}
    \end{gathered}
    \right|^2
\\
&+
     \left(
       \begin{gathered}
    \includegraphics[scale=0.8]{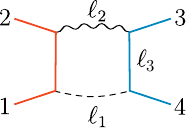} 
    \end{gathered}
    \right)^{\hspace{-0.2cm}2}
    +
\left(
    \begin{gathered}
        \includegraphics[scale=0.8]{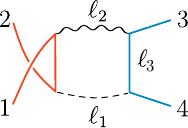}
    \end{gathered}
    \right)^{\hspace{-0.2cm}2}
     +
    \left(
    \begin{gathered}
        \includegraphics[scale=0.8]{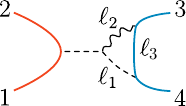}
    \end{gathered}
    \right)^{\hspace{-0.2cm}2}\\
      &
      +
    \left(
     \begin{gathered}
     \includegraphics[scale=0.8]{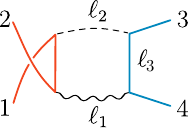}
     \end{gathered}
     \right)^{\hspace{-0.2cm}2}
      +
    \left(
    \begin{gathered}
        \includegraphics[scale=0.8]{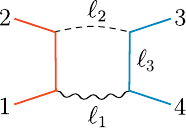}
    \end{gathered}
    \right)^{\hspace{-0.2cm}2}
     +
    \left(
    \begin{gathered}
        \includegraphics[scale=0.8]{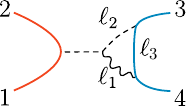}
    \end{gathered}
    \right)^{\hspace{-0.2cm}2},
    \end{aligned}
\end{equation}
where, again, only the labeled propagators $\ell_1$, $\ell_2$ and $\ell_3$ are on shell. We find that the numerators fixed by the gauge-theory cuts exactly recover the expression in \eqn{eq:NMC cut 1-loop gravity}.

\subsection{Bubble diagrams and cuts}

\noindent
As discussed in section \ref{Section:Why dilatons}, the double copy of YM leads not only to graviton and dilaton states, but involves also the axion (antisymmetric tensor $B^{\mu \nu}$ in general dimension). The axion cannot be sourced by massive scalars, dilatons or gravitons, and thus it can only appear in loops for amplitudes that have no external axions. For the one-loop four-point amplitude with external massive scalars, the axion can only appear in the massless bubble graph, which we now analyze in detail. 

\subsubsection*{Maximal cut}

While our current double-copy prescription does not include an axion ghost, we can work out the axion contribution by hand for any gravitational cut. Since the axion only appears in the massless bubble, we now consider the maximal cut of this topology. From the \ghost{} double copy, we have three contributions with a dilaton, to which we add an axion remainder,
\begin{equation}\label{Eq: Bubble max cut}
    \begin{aligned}
    \left.
        \begin{gathered}
            \includegraphics[scale=0.9]{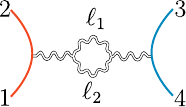}
        \end{gathered}
\right|_{\text{DC}+\text{axion}}
        \equiv& 
        \hspace{0.45cm}
        \left(
\begin{gathered}
\includegraphics[scale=0.9]{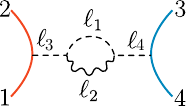}
\end{gathered}
\right)^{\hspace{-0.2cm}2}
+
\left(
\begin{gathered}
\includegraphics[scale=0.9]{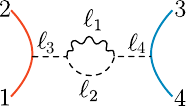}
\end{gathered}
\right)^{\hspace{-0.2cm}2}
\\
&+
\left(
 \begin{gathered}
\includegraphics[scale=0.9]{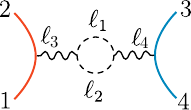}
\end{gathered}
\right)^{\hspace{-0.2cm}2}
+
\hspace{0.35cm}
\begin{gathered}
\includegraphics[scale=0.9]{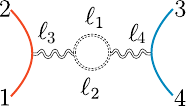}
\end{gathered},
    \end{aligned}
\end{equation}
where the last diagram with a dotted double line in the loop, denotes the axion contribution. The gauge-theory numerators that appear in this expression are given by the maximal cuts
\begin{equation}
\begin{gathered}
\includegraphics[scale=0.9]{figures/bubblemaxcut1.pdf}
\end{gathered}
= (k_1 \cdot \ell_2)(k_3 \cdot \ell_2)
\,
,
\hspace{0.7cm}
    \begin{gathered}
\includegraphics[scale=0.9]{figures/bubblemaxcut2.pdf}
\end{gathered} 
= 0\,
,
\end{equation}
where we recall that we use a single \ghost{} state in the loop. 

To work out the axion remainder of this cut, we can compare with the gravitational difference cut, 
\begin{equation}
\begin{aligned}
\left.
    \begin{gathered}
     \includegraphics{figures/bubblemaxcut1_gravity_thick.pdf}
     \end{gathered}
     \right|_{\Delta \mathfrak{C}}
     =
     \left( 
\frac{1}{2}(D_s-3)(D_s-2) + 1
     \right)
     \left(k_1\cdot \ell_2\right)^2 \left(k_3\cdot \ell_2\right)^2.
\end{aligned}
\end{equation}
Thus we see that the term proportional to $\frac{1}{2}(D_s-3)(D_s-2)$, i.e. the number of $B^{\mu\nu}$ states in $D_s$ dimensions, must be attributed to the axion remainder.

\subsubsection*{Two-particle cut (N${}^2$MC)}

We have verified our double-copy construction on all the maximal cuts and as well as on non-singular NMCs, with the minor caveat that we need to add or subtract axion contributions by hand. The final check is to perform the physical two-particle cuts, which will probe non-trivial contact terms associated to the dilaton and axion. The only two-particle cut that is interesting to discuss in detail, is the one where the intermediate on-shell legs are massless particles. The graphs that contribute to this two-particle cut are
\begin{equation}\label{Eq: Gravity n2-max cut}
\begin{aligned}
\mathfrak{C}_{\text{DC}+\text{axion}}^{\text{N${}^2$MC}}=&
\begin{gathered}
    \includegraphics[scale=0.9]{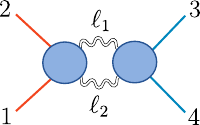}
    \end{gathered}
    \\
    =& 
    \hspace{0.45cm}
    \begin{gathered}
     \includegraphics[scale=0.9]{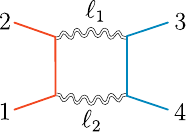}
     \end{gathered}
     +
      \begin{gathered}
     \includegraphics[scale=0.9]{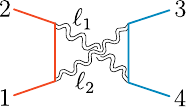}
     \end{gathered}
     +
      \begin{gathered}
     \includegraphics[scale=0.9]{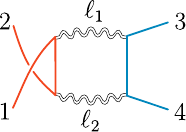}
     \end{gathered}
     \\
     &+
     \begin{gathered}
     \includegraphics[scale=0.9]{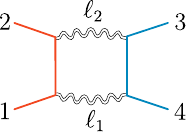}
     \end{gathered}
     +
      \begin{gathered}
     \includegraphics[scale=0.9]{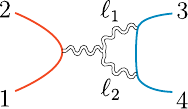}
     \end{gathered}
     +
      \begin{gathered}
     \includegraphics[scale=0.9]{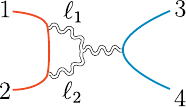}
     \end{gathered}
     \\
     &+
     \begin{gathered}
     \includegraphics[scale=0.9]{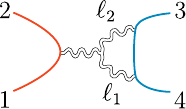}
     \end{gathered}
     +
      \begin{gathered}
     \includegraphics[scale=0.9]{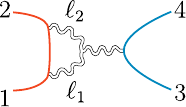}
     \end{gathered}
     +
      \begin{gathered}
     \includegraphics[scale=0.9]{figures/bubblemaxcut1_gravity_thick.pdf}
     \end{gathered}
     \\
     &+
\begin{gathered}
\includegraphics[scale=0.9]{figures/bubblemaxcut4.pdf}
\end{gathered},
    \end{aligned}
\end{equation}
While the first nine graphs are obtained from the double copy, we only schematically indicate the topologies. Drawing out every individual contribution, which has at least one internal dilaton line, would result in 26 graphs, which we refrain from doing. Comparing the double copy of the graphs determined \eqn{Eq: Gravity n2-max cut} to the gravity difference cut -- an expression which is too big to display here -- we find that they match up to terms that can all be attributed to the massless bubble. At first sight, these mismatching terms could be further attributed to the axion remainder, the graph on the last line of \eqn{Eq: Gravity n2-max cut}; however, we find this to not quite work out.  We now take a closer look at these bubble terms and untangle their origin.

First, we need to work out the axion remainder starting from the naive double copy of YM on the two-particle cut. This is given by applying the axion state projectors on the tensor indices of the internal states associated to the loop lines $\ell_1$ and $\ell_2$, 
\begin{align}
\begin{gathered}
    \includegraphics[scale=0.9]{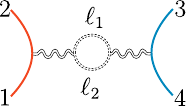}
\end{gathered}
\equiv & \ 
M_{\rm YM^2}(\r1,\r2,\ell_1,\ell_2) \Pi_{\rm axion}(\ell_1)\Pi_{\rm axion}(\ell_2) M_{\rm YM^2}(-\ell_2,-\ell_1,\b3,\b4)
\nn\\
=&
\hspace{0.3cm}
\frac{1}{4}(D_s-2) (D_s-3)\frac{ 2 (k_3\cdot \ell)^2(k_1\cdot \ell )^2+s_{12}(k_3\cdot \ell )(k_1\cdot \ell  ) (k_3-k_1)\cdot \ell }{s_{12}^2}
\nn\\
&-\frac{1}{8}(D_s-4)(D_s-5) (k_1\cdot \ell )( k_3\cdot \ell )
+ 2(D_s-4)\frac{ t_{24} (k_1\cdot \ell )( k_3\cdot \ell )}{s_{12}}
\nn\\
&+\frac{1}{8} (D_s-4) \left(\left(k_1\cdot \ell -t_{24}\right)^2+\left(k_3\cdot \ell +t_{24}\right)^2 +2 t_{24} (k_3-k_1)\cdot \ell \right)
\nn\\
&+\frac{1}{8}(D_s-5) \frac{ \left(s_{12}-4 \mu _3^2\right) \left(k_1\cdot \ell \right){}^2+ \left(s_{12}-4 \mu _1^2\right) (k_3\cdot \ell )^2
}{ s_{12}}
\nn\\&
+\frac{1}{4}(D_s-5) 
\left( \mu _1^2 (k_3\cdot \ell )-2 \mu _3^2 (k_1\cdot \ell )\right)
\nn\\
&+\frac{1}{4} \left(\mu _1^2 \mu _3^2-k_1\cdot \ell  k_3\cdot \ell \right) - \frac{1}{8}  ((k_1 \cdot \ell)^2+ (k_3 \cdot \ell)^2),
    \end{align}
where $\ell = \ell_1$, and $ \Pi_{\rm axion}^{\mu \nu;\rho \sigma}(\ell_i)=(P^{\mu \rho}P^{\nu\sigma}-P^{\mu \sigma}P^{\nu\rho})/2$, with vector state projector $P^{\mu \nu}=\eta^{\mu \nu}-2\ell_i^{(\mu} q^{\nu)}/\ell_i\cdot q$. 

Note that the axion contribution to ${\rm YM}{}^2$ should vanish in $D\le 3$, which we can make manifest by writing it in terms of Gram determinants. Indeed, the above expression for the axion remainder can be simplified to
\begin{equation}
\begin{aligned}
\begin{gathered}
    \includegraphics[scale=0.9]{figures/bubblecut_1_axion.pdf}
\end{gathered}
=&\hspace{0.3cm} \frac{(D_s - 2)(D_s- 3)}{2 s_{12}^2}  G\left(
\begin{matrix} 
\ell&k_1 \\
\ell& k_2
\end{matrix}\right)
G\left(
\begin{matrix}
\ell&k_3 \\
\ell& k_4
\end{matrix}\right)  
+ \frac{(D_s - 4)}{s_{12}^2} G(\ell, k_1, k_2, k_3)  \\
&-(D_s - 3)\frac{\mu^2_1~ G(\ell,k_3,k_4) + \mu^2_3 ~G(\ell, k_1, k_2)}{2 s_{12}^2}\,,
    \end{aligned}
\end{equation}
where $G(v_i)=G(v_1,v_2,\ldots,v_n)={\rm Det}(v_i\cdot v_j)$, and $G({}^{v_i}_{w_j})={\rm Det}(v_i\cdot w_j)$. This makes it manifest that it vanishes in $D=D_s=1,2,3$, thus verifying the expression. We note that on this form, the individual terms can be given physical interpretations.\footnote{The physical interpretation of the different terms should be helpful for constructing axion ghosts in a double copy that only involves bosonic fields (in contrast to the fermionic construction of \rcite{Johansson:2014zca}). For example, one-loop terms proportional to $(D_s-2)(D_s-3)$ always come from effective scalar modes, equivalent to the \ghost{} contribution.} The first term exactly matches the bubble dilaton contribution in $D_s=4$. Since this term scale as $\sim D_s^2$ in the infinite-$D_s$ limit the contribution corresponds to interactions where the axion polarizations never talk to the momenta or other states, hence the interactions are the same as for a scalar. The third term has the interpretation that one of the $s$-channel particles is a dilaton (the other one is a graviton/dilaton), hence it couples to the mass $\mu_i^2$. The second term vanishes in $D_s=4$, hence this contribution comes from a genuine tensor field $B^{\mu\nu}$ and not the pseudo-scalar realization of the axion. 

Finally, we circle back to the main check for the two-particle cut: comparing the gravitational difference cut and the \ghost{} double copy. After taking into account the axion remainder, we find that there is yet another small remainder that is not accounted for in our construction. All of the pole terms and most of the contact terms in the cut cancel out once adding the axion remainder; however, the remaining unaccounted piece is
\begin{align}
\Delta &=\mathfrak{C}^{\text{N${}^2$MC}}_{\text{YM}^2}(\r1,\r2,\b3,\b4;\ell)  -  \mathfrak{C}^{\text{N${}^2$MC}}_{\text{GR}}(\r1,\r2,\b3,\b4;\ell)
-
\mathfrak{C}^{\text{N${}^2$MC}}_{\text{DC}+\text{axion}}(\r1,\r2,\b3,\b4;\ell)\nn\\
&=
-\frac{1}{2^7}\Big(s_{12}^2-\big(2\ell\cdot(k_1-k_2)\big)^2-\big(2\ell\cdot(k_3-k_4)\big)^2 \Big).
\end{align}
While this remainder is undesirable, and suggest that \ghost{} double copy needs a modification at loop level, one may also take the perspective that this simple contribution can be subtracted out by hand as we did for the axion. Until we have a complete double copy for both the dilaton and axion ghost, we need not worry about it. Furthermore, if we are interested in classical scattering of black holes, this contribution is a quantum effect that we can ignore. 
See appendix~\ref{Appendix: loop}, where this discrepancy is analyzed from a slightly different perspective.

\section{Conclusion}

The double copy \cite{Bern:2008qj,Bern:2010ue} is a highly efficient method for computing scattering amplitudes in a plethora of gravitational theories~\cite{Bern:2022wqg,Adamo:2022dcm,Bern:2023zkg,Bern:2011rj,Chiodaroli:2014xia,Chiodaroli:2015rdg,Chiodaroli:2015wal,Chiodaroli:2017ehv,Johansson:2017bfl,Johansson:2017srf,Johansson:2018ues,Azevedo:2018dgo,Chiodaroli:2018dbu,Chiodaroli:2021eug,Chiodaroli:2023tvo,Johansson:2023ymb} from gauge theory input. However, it remains an open problem how to best engineer the desired particle spectrum and interactions in certain theories. The most pressing example is general relativity coupled to massive matter, where the naive YM double copy sources unwanted massless states: the dilaton and axion (or $B^{\mu\nu}$ field in general dimension). The dilaton and axion can be removed by hand, using on-shell projectors \cite{Bern:2019crd,Carrasco:2020ywq,Bern:2021dqo,Carrasco:2021bmu,Bern:2024vqs}, or equivalently, by explicitly carrying out state sums corresponding to selecting only gravitons~\cite{Brandhuber:2021kpo,Brandhuber:2021eyq,Brandhuber:2021bsf}, in appropriate unitarity cuts. The combination of the double copy and projection in \rcite{Carrasco:2021bmu} provides a powerful and robust tool for calculating GR amplitudes  which eliminates the need for ans\"atze in gravitational and multiloop amplitudes, and indeed supplies the necessary data used to fix the amplitudes in this paper. However, such approaches cannot be applied to off-shell intermediate states, and thus one loses some of the powerful relations from color-kinematics duality. Furthermore, the explicit removal of unwanted terms may spoil the nice feature that the gravitational contributions are sums of squares.  A method that goes hand-in-hand with color-kinematics duality and the double copy is to introduce new gauge-theory states that behave as ghosts in the double copy, such that they automatically cancel out the unwanted states. 

Such double-copy ghosts were first introduced in~\rcite{Johansson:2014zca} and solved the problem of obtaining pure-GR (graviton-only) amplitudes in four dimensions, using fermionic fields in the underlying gauge theory. Since the fermionic fields have to form closed loops, they are less well suited for removing dilatons and axions that are linearly coupled to massive matter. Instead, bosonic fields may be used as ghosts in this case, as was first discussed in \rcite{Luna:2017dtq}, where it was observed that a scalar ghost can be used to remove the dilaton in simple four- and five-point tree-level graphs. For tree-level double-copy constructions that only involve spinless massive matter, we can ignore the axion since it does not couple linearly to matter. At loop level, the axion will of course contribute since it can form closed loops. It then likely needs to be treated along the lines of~\rcite{Johansson:2014zca} where the underlying gauge theories are supplemented by fermions or other spinning ghost states. 

In this paper, we restrict our attention to the problem of removing dilatons from the double copy in general spacetime dimension. Modulo further checks, this gives a prescription for obtaining tree-level GR amplitudes with massive scalar matter, and it provides a stepping stone towards solving the corresponding loop-level problem. We use several complementary methods for working out the states and interactions in the underlying gauge theory with \ghost{} ghost states, such that the dilaton contaminations cancel out from GR. First, we introduce the massive scalar states and \ghost{} via Kaluza-Klein compactification (and truncation), where the latter state is identified with the higher-dimensional component of the gluon. We check that the three-point Kaluza-Klein amplitudes are the needed ones such that the \ghost{} double copy matches those coming from the dilaton. Second, we work out the gauge-theory Lagrangian where all interactions but one are determined from the compactification. Only the two-scalar-two-\ghost{} interaction is modified by a non-trivial factor $\fD$ that we anticipate from a six-point matching calculation to GR. 

As a third main approach, we bootstrap higher-point gauge-theory \ghost{} amplitudes such that they satisfy color-kinematics duality and reproduce the unwanted dilaton-mediated contributions coming from the naive YM double copy. We consider a variety of external states at four points, analyze the massive six-point amplitude at tree level, and finish by computing the color-kinematics-satisfying numerators of the one-loop four-scalar amplitude. We find that the \ghost{} interactions are compatible both with necessary kinematic Jacobi relations from color-kinematic duality and optional bonus relations, and we introduce a new type of bonus relation that relates numerators with different numbers of \ghost{} lines. This relation involves a multiplicative factor $\fD$, that takes into account the dilaton normalization factors $(D_s-2)$ that differ between the graphs. The \ghost{} double copy is asymmetric in the sense that certain numerators depend on the $\fD$ factor with opposite sign choices for the two copies. We confirm that the construction works for tree amplitudes with up to six massive external scalars, such that the \ghost{} double copy removes the unwanted dilaton states in GR amplitudes. Further checks of the construction is needed, such as amplitudes with eight massive external scalars. We leave this to future work.  

The one-loop four-scalar amplitude is worked out in detail so that its double copy matches the dilaton exchange contributions from all gravitational graphs except for the massless bubble. The massless bubble graph has contributions from an axion (or $B^{\mu\nu}$ field) in the loop, which we work out by hand. After accounting for the axion bubble, we find that the two-particle cut still gives a simple local bubble remainder that is not accounted for in our double-copy prescription. While the current double-copy approach needs to be supplemented by an axion ghost before it becomes ripe for one-loop applications, the discrepancy in the massless bubble needs to be better understood. One may speculate that it can be resolved in a variety of possible ways: 1) additional dilaton ghost contributions may be needed for diagrams where the dilaton can propagate in a loop; 2) the appearance of the $\fD$ factor in the bonus relation and in contact interactions needs a more sophisticated treatment where the value of $\fD$ may depend on the details of the graph; 3) the \ghost{} may need to be promoted to a complex field so that graph automorphisms can be relaxed giving more freedom for the kinematic numerators; 4) consider the option of using a vector field as ghost, since canceling the $B^{\mu\nu}$ contributions in general dimension necessarily requires a field with Lorentz indices. We leave this as an open problem for future work, especially in the context of computing higher-loop amplitudes. 

As a final remark, we note that it would be interesting to generalize the work in this paper to double copies with spinning massive matter. It is known that the double copy works well~\cite{Johansson:2019dnu,Bautista:2019evw,Chiodaroli:2021eug,Cangemi:2023ysz} for generating tree-level amplitudes~\cite{Arkani-Hamed:2017jhn,Guevara:2018wpp,Chung:2018kqs,Arkani-Hamed:2019ymq} that describe a spin $s\le2$ Kerr black hole that interacts with any number of gravitons. For tree amplitudes with two or more Kerr black holes (or spinning massive matter sources), the dilaton and axion mediate long range forces and need to be removed from the naive double copy. While this can be done using on-shell graviton projectors in unitary cuts~\cite{Bern:2019crd,Carrasco:2020ywq,Bern:2021dqo,Carrasco:2021bmu}, it would be conceptually interesting if this can also be achieved in a harmonized double-copy framework that uses ghost states.  

\section{Acknowledgments}

We thank Alex Edison, Andres Luna, Kays Haddad, Alex Ochirov, Donal O'Connell and Paolo Pichini for useful discussions and related collaborations. We particularly want to thank John Joseph Carrasco for collaboration at an early stage of this project. The research is supported by the Knut and Alice Wallenberg Foundation under grants KAW 2018.0116 (From Scattering Amplitudes to Gravitational Waves) and KAW 2018.0162 (Exploring a Web of Gravitational Theories through Gauge-Theory Methods).
This research was supported in part by grant NSF PHY-2309135 to the Kavli Institute for Theoretical Physics (KITP). H.J. thanks KITP for the hospitality during the completion of this work.

\appendix
\section{Alternative analysis of one-loop numerators}\label{Appendix: loop}

Consider the two-particle cut of the one-loop amplitude. The two-dilaton exchange coming purely from the double copy of gluons $H_{\mu \nu} \sim A_\mu \otimes A_\nu$, is given by the cut
\begin{align}
\begin{gathered}
\includegraphics[scale=1]{figures/n2cut1_thick.pdf}
\end{gathered}
&=
\mathfrak{C}_{HH}:=M(\r1, \r2, \ell_1, \ell_2) \Pi_\text{dilaton}(\ell_1)  \Pi_\text{dilaton}(\ell_2) M( -\ell_2,-\ell_1, \b3, \b4) \nn \\
&
= \frac{1}{(D_s{-}2)^2}\Big(\frac{{\cal N}_t}{t_{2\ell}} + \frac{{\cal N}_u}{t_{1\ell}} + (D_s{-}2)\frac{{\cal N}_s}{s_{12}} \Big)\Big(\frac{ {\cal N}_{\tilde t}}{\tilde t_{3\ell}} + \frac{ {\cal N}_{\tilde u}}{\tilde t_{4\ell}} + (D_s{-}2) \frac{ {\cal N}_{\tilde s}}{s_{34}} \Big)\,, 
\end{align}
where $M(\r1, \r2, \ell_1, \ell_2)=s_{12}\frac{t_{2\ell}}{t_{1\ell}}\big[A(\r1, \r2, \ell_1, \ell_2)\big]^2$ are double copies of YM scalar/gluon tree amplitudes, $\Pi^{\mu \nu;\rho\sigma}_\text{dilaton}=\frac{1}{D_s-2}P^{\mu\nu}P^{\rho\sigma}$ is the dilaton projector, and we use the shorthands $t_{i\ell}=2 k_i \cdot \ell_1= - \tilde t_{i\ell}$. Since the gauge-theory amplitudes are unique by gauge invariance, the gravitational numerators are straightforward to compute
\begin{align}
{\cal N}_t= \frac{1}{2}m^2 (2m^2 +t_{2\ell}) \,,~~~{\cal N}_u=  \frac{1}{2} m^2 (2m^2 + t_{1\ell})\,,~~~{\cal N}_s= -  \frac{1}{4}t_{1\ell} t_{2\ell}\,.
\end{align}
The tilde numerators are given by the same expressions, except that $t_{1\ell}\to \tilde t_{4\ell}$, $t_{2\ell} \to \tilde t_{3\ell}$ and $m \to \tilde m$. 

Likewise, the two-dilaton exchange coming from the double copy of a \ghost{} exchange is
\begin{align}
\mathfrak{C}_{H\varphi}:= &
\begin{gathered}
    \includegraphics[scale=1.0]{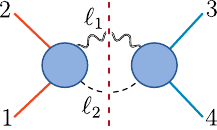}
\end{gathered}
\\
=
&M(\r1, \r2, \ell_1,\g{\ell_2}) \Pi_\text{dilaton}(\ell_1)  M(-\g{\ell_2}, -\ell_1, \b3,\b4) \nn \\
=& \frac{1}{(D_s-2)^2}\Big(\frac{{\cal N}_t}{t_{2\ell}} +\frac{{\cal N}_u}{t_{1\ell}} \Big)\Big(\frac{ {\cal N}_{\tilde t}}{\tilde t_{3\ell}} + \frac{ {\cal N}_{\tilde u}}{\tilde t_{4\ell}}\Big) \,,
\end{align}
where $M(\r1, \r2, \ell_1, \g{\ell_2})=s_{12}\frac{t_{2\ell}}{t_{1\ell}}\big[A(\r1, \r2, \ell_1, \g{\ell_2})\big]^2$ are double copies tree amplitudes involving scalars, a gluon and a \ghost. Again these gauge-theory amplitudes are unique, and thus the gravity numerators are obtained by straightforward computation
\begin{align}
{\cal N}_t= \frac{1}{2} m^2 (2m^2 + t_{2\ell}) \,,~~~{\cal N}_u=  \frac{1}{2} m^2 (2m^2 + t_{1\ell})\,,
\end{align}
Note that now there is no $s$-channel contribution, since that would correspond to a forbidden cubic-\ghost{} vertex. Note that if we swap legs $\ell_1$ and $\ell_2$ we get the same expression, hence $\mathfrak{C}_{\varphi H}= \mathfrak{C}_{H\varphi}$.

Finally the pure two-\ghost\, exchange contributions is given by
\begin{align}
\mathfrak{C}_{\varphi \varphi}:=&\begin{gathered}
    \includegraphics[scale=1.0]{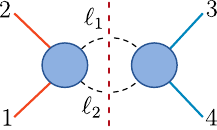}
\end{gathered}\\
=& \frac{1}{(D_s-2)^2}\Big(\frac{{\cal N}_t}{t_{2\ell}} +\frac{{\cal N}_u}{t_{1\ell}} \Big)\Big(\frac{ {\cal N}_{\tilde t}}{\tilde t_{3\ell}} + \frac{ {\cal N}_{\tilde u}}{\tilde t_{4\ell}}\Big) \nn \\
&- \frac{1}{(D_s-2)}\Big(\frac{{\cal N}_t}{t_{2\ell}} +\frac{{\cal N}_u}{t_{1\ell}} \Big)\frac{ {\cal N}_{\tilde s}}{s_{34}}
 \nn \\
 &- \frac{1}{(D_s-2)}\Big(\frac{{\cal N}_{\tilde t}}{\tilde t_{3\ell}} +\frac{{\cal N}_{\tilde u}}{\tilde t_{4\ell}} \Big) \frac{ {\cal N}_{s}}{s_{12}}  \nn \\
&-\frac{{\cal N}_{s}}{s_{12}} \frac{{\cal N}_{\tilde s}}{s_{34}} \,,
\end{align}
where the numerators are now unknown. 
We can solve for these unknows, by imposing the equation that the signed sum of cut contributions vanish,
\begin{align}
\mathfrak{C}_{HH}-\mathfrak{C}_{\varphi H}-\mathfrak{C}_{H\varphi}+\mathfrak{C}_{\varphi \varphi}=0\,.
\end{align}
The solution is unique, given that we impose the correct Bose symmetries of the diagrams, and the numerators coincide with the above ones
\begin{align}
{\cal N}_t= \frac{1}{2} m^2 (2m^2 + t_{2\ell}) \,,~~~{\cal N}_u=  \frac{1}{2} m^2 (2m^2 + t_{1\ell})\,,~~~{\cal N}_s= - \frac{1}{4} t_{1\ell} t_{2\ell}\,.
\end{align}
Since these are gravitational numerators, one would need to factorize them in order to get gauge-theory numerators. There appears to be no elegant way to take a square root. However, an asymmetric factorization compatible with kinematic Jacobi relations, but different than \eqns{Eq: 2d2m final numerator}{Eq: 2d2m second final numerator}, seems possible:
\begin{align}
&N_i~\text{numerators:}~~~~~ \Big\{m^2,~~ -m^2- \frac{1}{2}t_{1\ell},~~ \frac{1}{2}t_{1\ell}\Big\}\,, \nn \\
&\tilde N_i~\text{numerators:}~~~~~ \Big\{m^2+ \frac{1}{2}t_{2\ell},~~ -m^2,~~ - \frac{1}{2}t_{2\ell}\Big\}\,,
\end{align}
where the sum of each numerator triplet is zero! However, since the $s$-channel numerator, $N_s=t_{1\ell}$,  is no longer asymmetric under exchange of $1\leftrightarrow 2$ nor $\ell_1\leftrightarrow \ell_2$, this implies that this can at best be made consistent if we promote both the \ghost{} and the massive fields to be complex scalars. This would alter the construction significantly, and likely introduce other issues because the action of sewing together \ghost{} states would be subject to new rules that respects the charge of these fields. A preliminary analysis suggests that this will lead to double counting of \ghost{} states, and thus over-cancellations of dilatons. We leave further analysis of this complex \ghost{} scenario to future work.

\bibliographystyle{JHEP}
\bibliography{dilatonbootstrap.bib}

\end{document}